\documentclass[11pt]{article}%
\usepackage{makeidx, siunitx, ragged2e}
\usepackage{booktabs,color}
\usepackage{amsfonts,bm}
\usepackage{amsmath}
\usepackage{amssymb}
\usepackage{graphicx}
\usepackage{rotating}
\usepackage{multirow}%
\usepackage{geometry,setspace}
\usepackage{natbib}
\usepackage{multicol}
\usepackage{subfigure}
\usepackage{rotating}
\usepackage{tikz}
\usepackage{pdfpages}
\usepackage{ulem}
\usepackage{booktabs}
\usepackage{lscape}
\usepackage{caption}
\usepackage{lipsum}
\usepackage{longtable}
\newcommand{\E}{\mathbb{E}}

\begin{document}

\title{A Natural Disasters Index}
\author{
Thilini V. Mahanama\thanks{Texas Tech University, Department of Mathematics
\& Statistics, Lubbock TX 79409-1042, U.S.A., thilini.v.mahanama@ttu.edu (Corresponding
Author).}
\and 
Abootaleb Shirvani\thanks{Texas Tech University, Department of Mathematics
\& Statistics, Lubbock TX 79409-1042, U.S.A., abootaleb.shirvani@ttu.edu.}
 }
\date{}
\maketitle

\begin{abstract}
Natural disasters, such as tornadoes, floods, and wildfire pose risks to life and property, requiring the intervention of insurance corporations. One of the most visible consequences of changing climate is an increase in the intensity and frequency of extreme weather events. The relative strengths of these disasters are far beyond the habitual seasonal maxima, often resulting in subsequent increases in property losses. Thus, insurance policies should be modified to endure increasingly volatile catastrophic weather events. We propose a Natural Disasters Index (NDI) for the property losses caused by natural disasters in the United States based on the ``Storm Data" published by the National Oceanic and Atmospheric Administration. The proposed NDI is an attempt to construct a financial instrument for hedging the intrinsic risk. The NDI is intended to forecast the degree of future risk that could forewarn the insurers and corporations allowing them to transfer insurance risk to capital market investors. This index could also be modified to other regions and countries.
\end{abstract}



\section{Introduction}\label{sec:intro}

Natural disasters are low-probability, high-consequence events that wreak havoc on financial security \citep{roth1998paying}.
The National Centers for Environmental Information (NCEI) reports the United States has experienced 69 natural disasters with losses exceeding one billion dollars between 2015 and 2019.
The accumulated loss exceeds \$535 billion at an average of \$107.1 billion/year. 
The trend of disaster frequency is expected to escalate over the years due to changes in climate which will result in deleterious losses \citep{lyubchich2017can}.
These volatile weather patterns will result in an inevitable challenge to the U.S.'s ability to sustain human and economic development \citep{NYTimes2017}.
As a result, weather risk markets need to be capable of offsetting the financial impacts of natural disasters \citep{varangis2003weather,dilley2005natural}.

The losses due to natural disasters exacerbate due to changes in population and national wealth density \citep{van1998united,bell2018changes}.
\cite{roth1998paying} suggest that if insurers are to retain profitability and solvency in the event of a major catastrophe that insurers must increase their prices for catastrophe insurance and reduce their exposure to risk.
Also, reinsurers undergo severe financial stress in facilitating catastrophe insurance by offering tenable reduction for risk in large catastrophic losses \citep{Lewis1996,liang2010optimal,zangue2016evaluating}.
However, the detrimental losses can be alleviated using protective measures such as preparedness, mitigation, and insurance \citep{kunreuther1996mitigating,ganderton2000buying}.
To better protect the clients, catastrophe insurance policies should ramp-up investments in cost-effective loss reduction mechanisms by better managing the risk.

According to \cite{barnett2007weather}, the weather index insurance can effectively transfer spatially covariate weather risks as it pays indemnities based on realizations of a weather index that is highly correlated with actual losses.
The securitization of losses from natural disasters provides a valuable novel source of diversification for investors.
Catastrophe risk bonds are a promising type of insurance-linked securities introduced to smooth transferring of catastrophic insurance risk from insurers and corporations to capital market investors by offering an alternative or complement of capital to the traditional reinsurance \citep{zangue2016evaluating}. 
\cite{cummins2004basis} describe three types of variables that pay off in insurance-linked securities: insurer-specific catastrophe losses, insurance-industry catastrophe loss indices, and parametric indices based on the physical characteristics of catastrophic events.

Unequivocally, the catastrophe losses and related risks inherent create uncertainty over the type of disaster event \citep{Lewis1996,BillionDollar}. 
For example, due to less coverage of insured assets and data latency in drought and flooding events, they tend to provide uncertain loss estimates compared to the losses of severe storm events in the United States \citep{BillionDollar,smith2015quantifying}. 
In consequence, prioritization for mitigating the risks can be diverse and complex. 

We propose a Natural Disasters Index (NDI) for the United States using the property losses reported in NOAA Storm Data \citep{StormData} between 1996 and 2018.
The NDI is aimed to assess the level of future systemic risk caused by natural disasters.
We follow the methods applied in \cite{trindade2020socioeconomic} on an ad hoc basis as a benchmark for NDI evaluation: (1) option pricing, (2) risk budgeting, and (3) stress testing.
We provide an evaluation framework for the NDI using a discrete-time generalized autoregressive conditional heteroskedasticity model to calculate the fair values of the NDI options.
Then, we simulate call and put option prices using the Monte Carlo method.
We distribute the cumulative risk attributed to our equally weighted portfolio into the risk contributions of each type of natural disaster.
Flood and flash flood are the main risk contributors in our portfolio according to our assessments using standard deviation and expected tail loss risk budgets.
Furthermore, we evaluate the portfolio risk of the NDI to mitigate risks using monthly maximum temperature and the Palmer Drought Severity Index (PDSI) as stressors.
We found the stress on maximum temperature significantly impacts the NDI compared to that of the PDSI at the highest stress level (1\%).

There have been similar attempts to develop indices in the past, though many of these are  no longer used because of inherent problems.
The first index-based catastrophe derivatives, CAT-futures, introduced by the Chicago Board of Trade using the ISO-Index was ineffective due to a lack of realistic models in the market \citep{christensen2000pricing}.
Secondly, the Property Claim Services (PCS) proposed the PCS-options based on the PCS-index.
\cite{biagini2008pricing} describe that the PCS-options slowed down due to market illiquidity.
Then, the New York Mercantile Exchange (NYMEX) designed catastrophe futures and options to enhance the transparency and liquidity of the capital markets to the insurance sector \citep{biagini2008pricing}.
\cite{kielholz1997insurance} further explain alternative risk transfer mechanisms within the context of natural catastrophe problems in the United States.

The proposed NDI attempts to address these shortcomings by creating a financial instrument for hedging the intrinsic risk induced by the property losses caused by natural disasters in the United States.
The vital objective of the NDI is to forecast the severity of future systemic risk attributed to natural disasters.
This provides advance warnings to the insurers and corporations allowing them to transfer insurance risk to capital market investors. 
Therefore, the proposed NDI will conspicuously help to make up the shortfall between the capital and insurance markets.
The NDI identifies the potential risk contributions of each natural disaster and provides options and futures.
Furthermore, the NDI could be modified to calculate the risk in other regions or countries using a data set comparable to NOAA Storm Data \cite{StormData}.

The contents of the rest of this paper are as follows. We provide an exploratory data analysis in section \ref{sec:data} before constructing the NDI.
Section \ref{sec:OP} presents the steps in option pricing and approximate call and put option prices for the NDI. 
In section \ref{sec:RB}, we provide standard deviation and expected tail loss risk budgets for natural disasters in the United States. 
We assess the performance of the NDI via a stress testing analysis in section \ref{sec:ST}. 
Finally, we make concluding remarks in section \ref{sec:DC}.

\section{Construction of the Natural Disasters Index (NDI)}\label{sec:data}

The National Oceanic and Atmospheric Administration (NOAA) has published information on severe weather events occurring in the United States between 1950 and 2018 in their ``Storm Data" database \citep{StormDataPrep}. 
We utilize the property losses caused by the following 50 types of natural disasters from 1996-2018 to construct a natural disasters index:

\begin{quote}
	Avalanche, Blizzard, Coastal Flood, Cold/Wind Chill, Debris Flow, Dense Fog, Dense Smoke, Drought, Dust Devil, Dust Storm, Excessive Heat, Extreme Cold/Wind Chill, Flash Flood, Flood, Frost/Freeze, Funnel Cloud, Freezing Fog, Hail, Heat, Heavy Rain, Heavy Snow, High Surf, High Wind, Hurricane (Typhoon), Ice Storm, Lake-Effect Snow, Lakeshore Flood, Lightning, Marine Dense Fog, Marine Heavy Freezing Spray, Marine High Wind, Marine Hurricane/Typhoon, Marine Lightning, Marine Strong Wind, Marine Thunderstorm Wind, Rip Current, Seiche, Sleet, Storm Surge/Tide, Strong Wind, Thunderstorm Wind, Tornado, Tropical Depression, Tropical Storm, Tsunami, Volcanic Ash, Waterspout, Wildfire, Winter Storm, Winter Weather.
\end{quote}

\begin{figure}[h!]
	\subfigure{\includegraphics[width=0.32\textwidth]{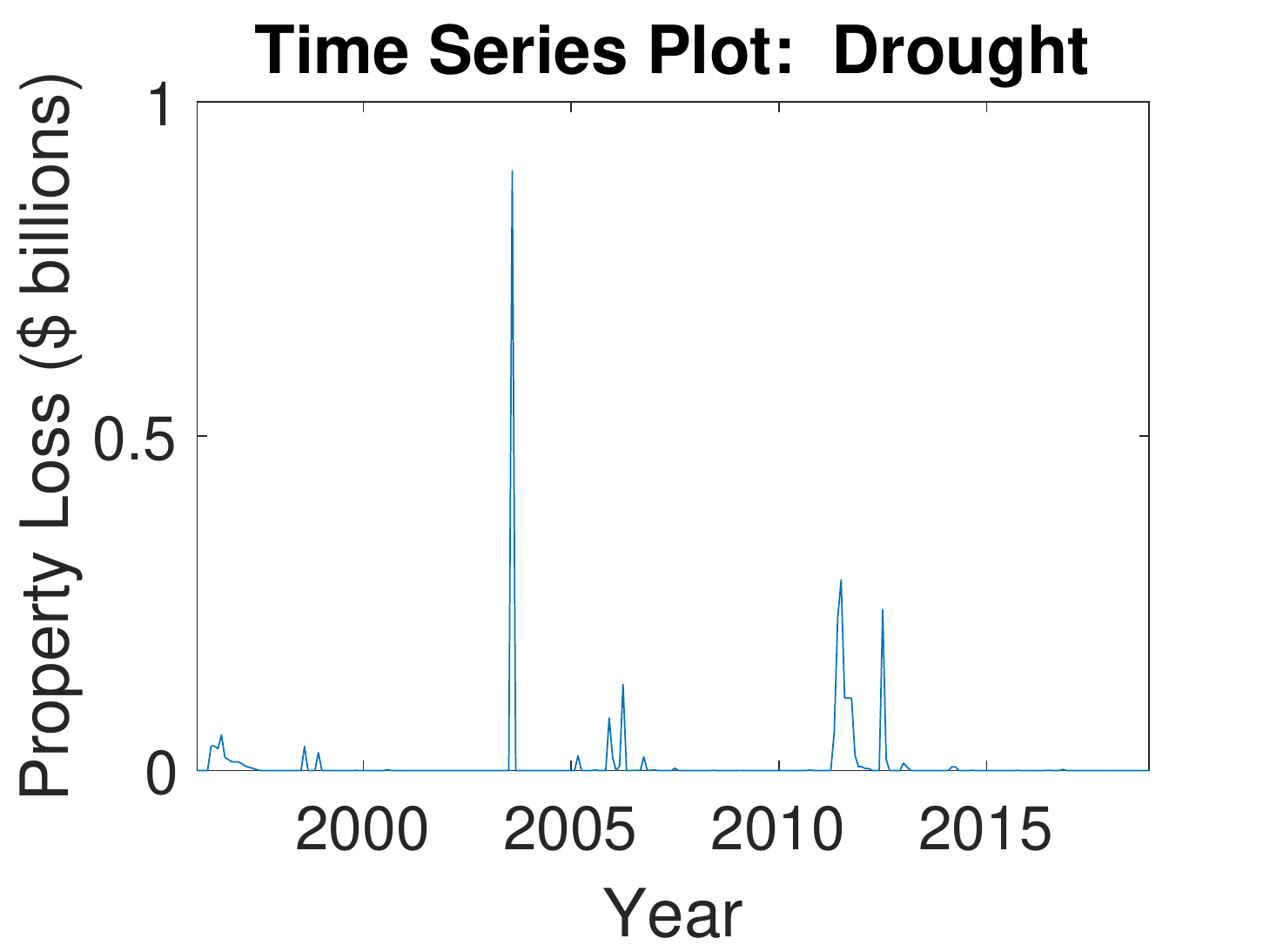}} 
	\subfigure{\includegraphics[width=0.32\textwidth]{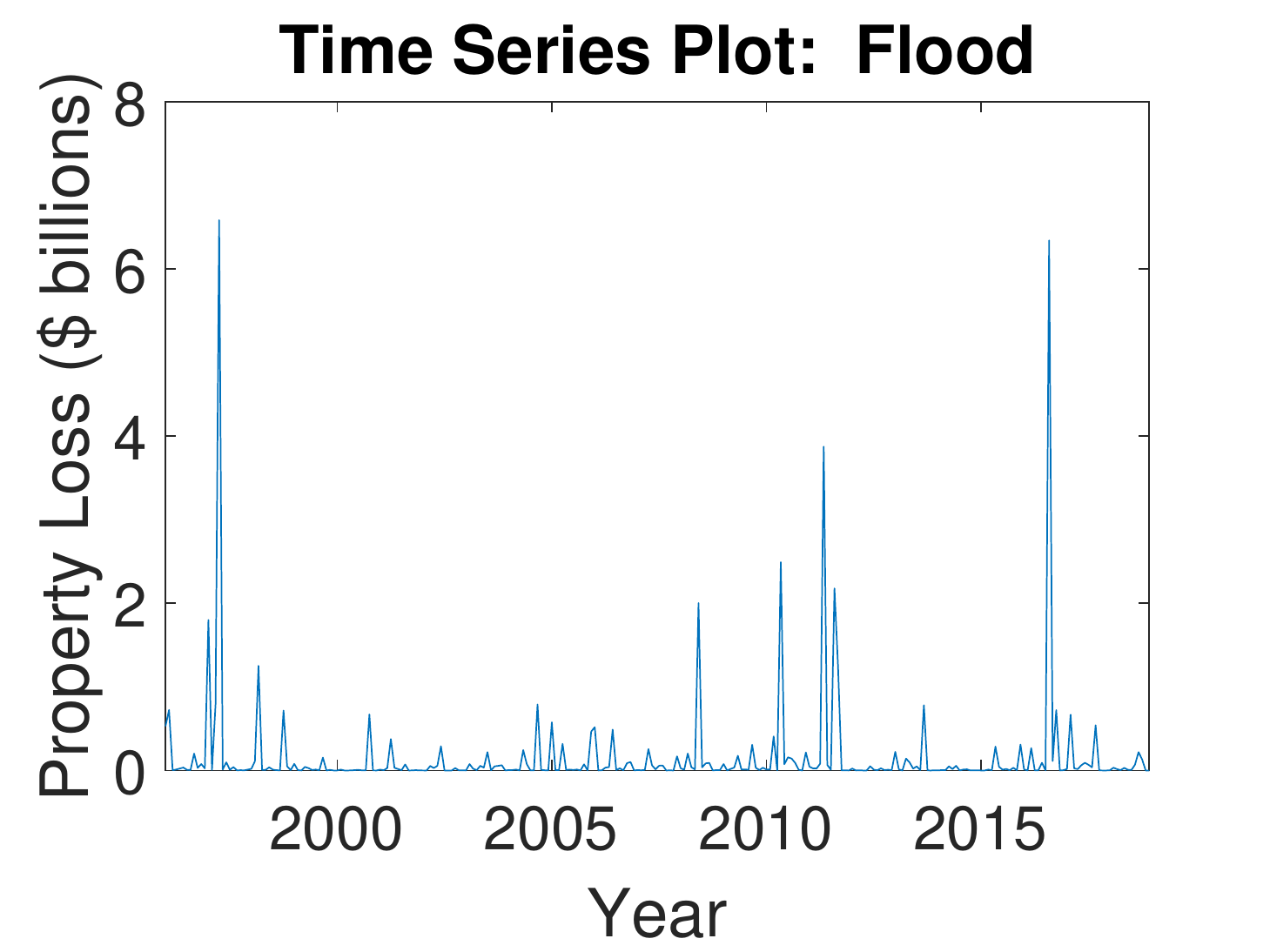}} 
	\subfigure{\includegraphics[width=0.32\textwidth]{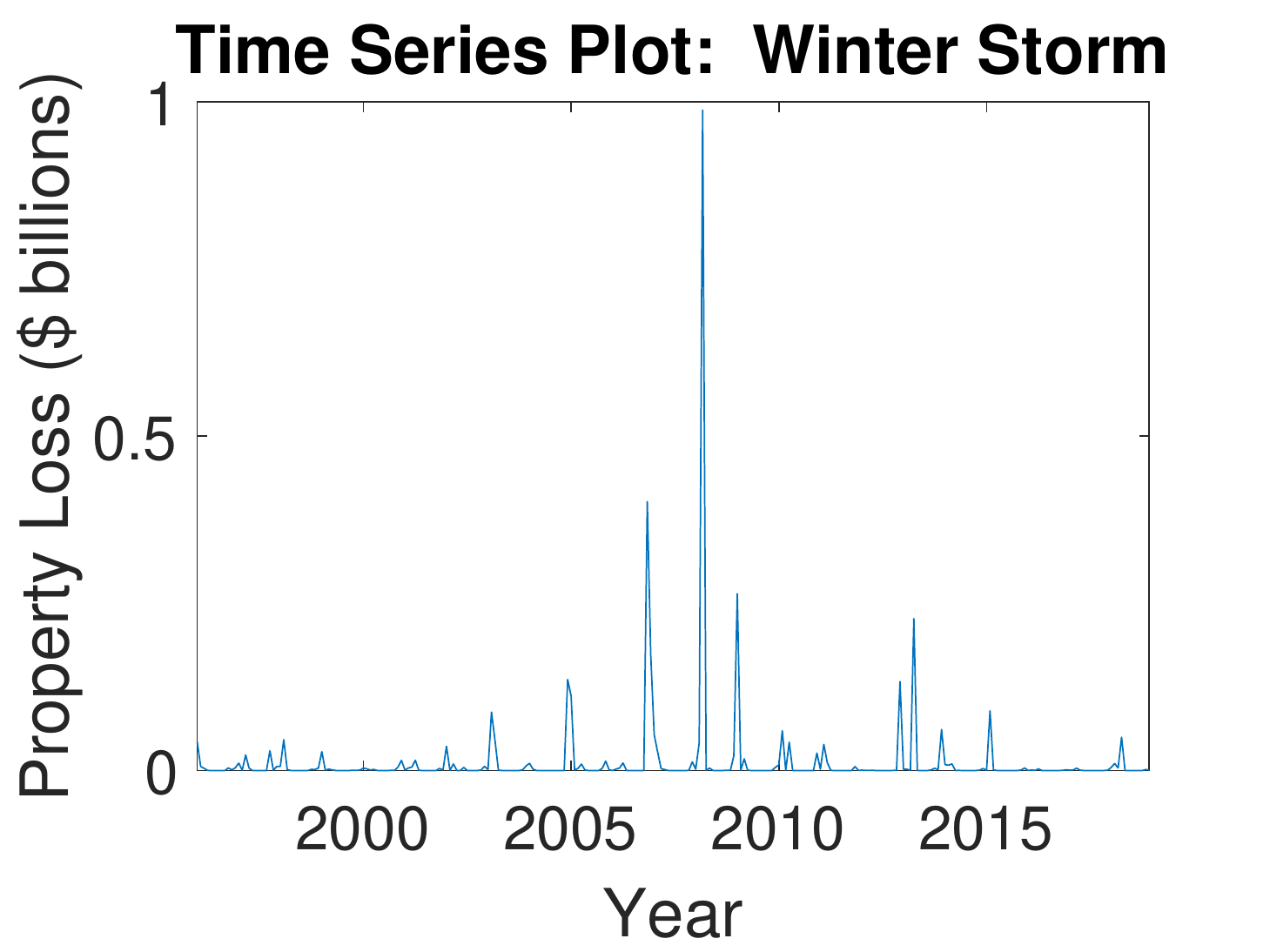}} 
	\subfigure{\includegraphics[width=0.32\textwidth]{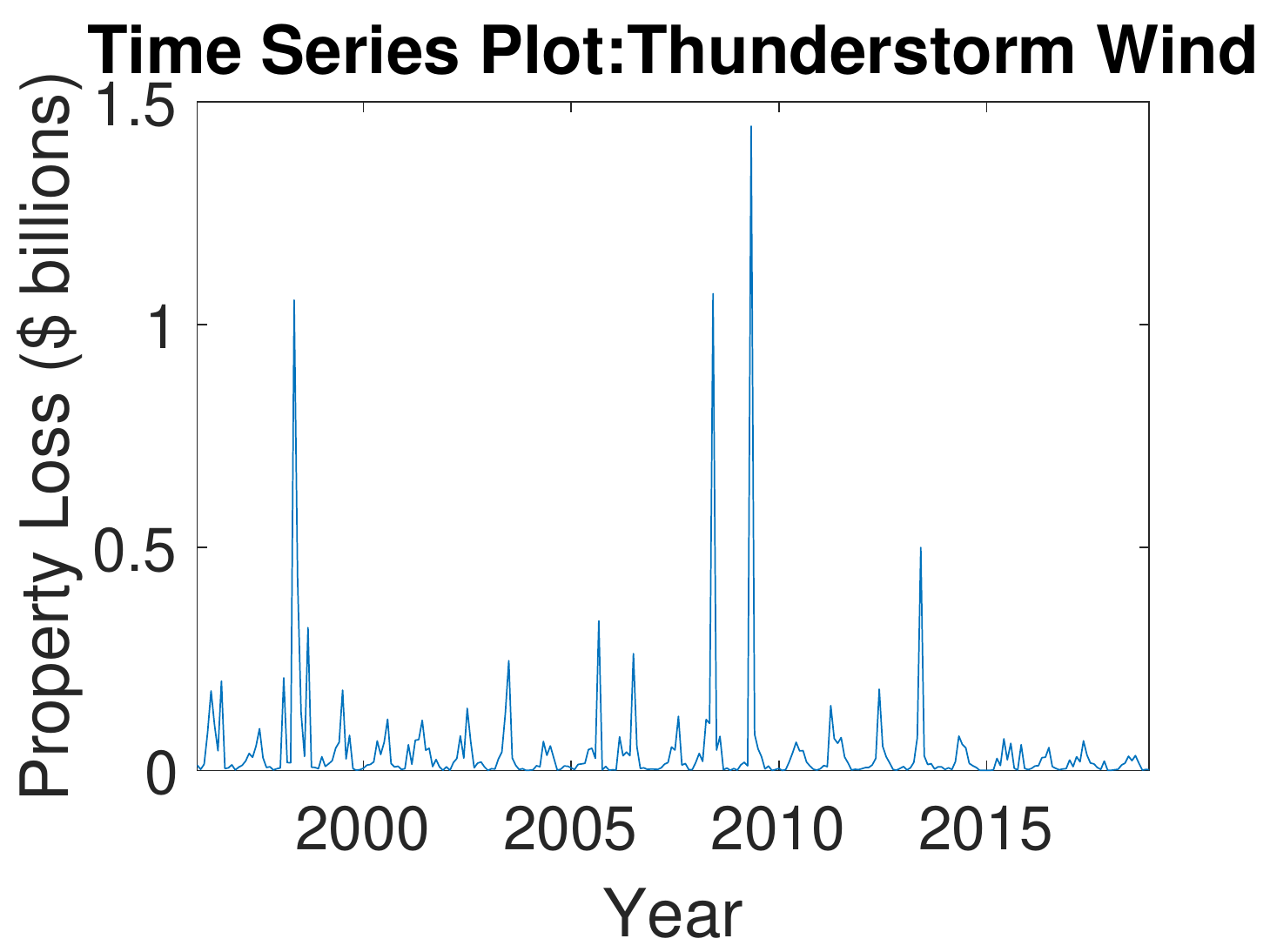}} 
	\subfigure{\includegraphics[width=0.33\textwidth]{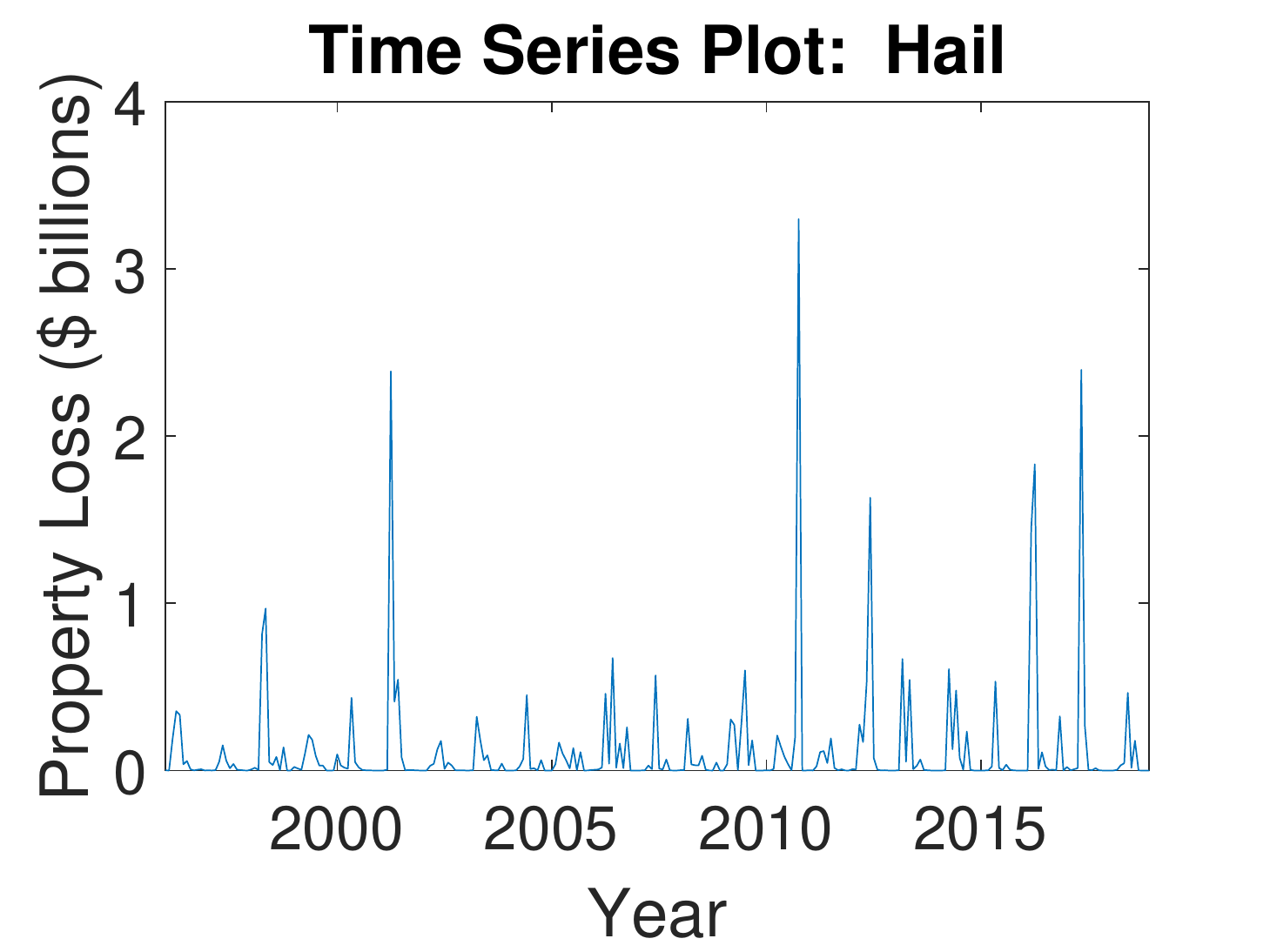}} 
	\subfigure{\includegraphics[width=0.33\textwidth]{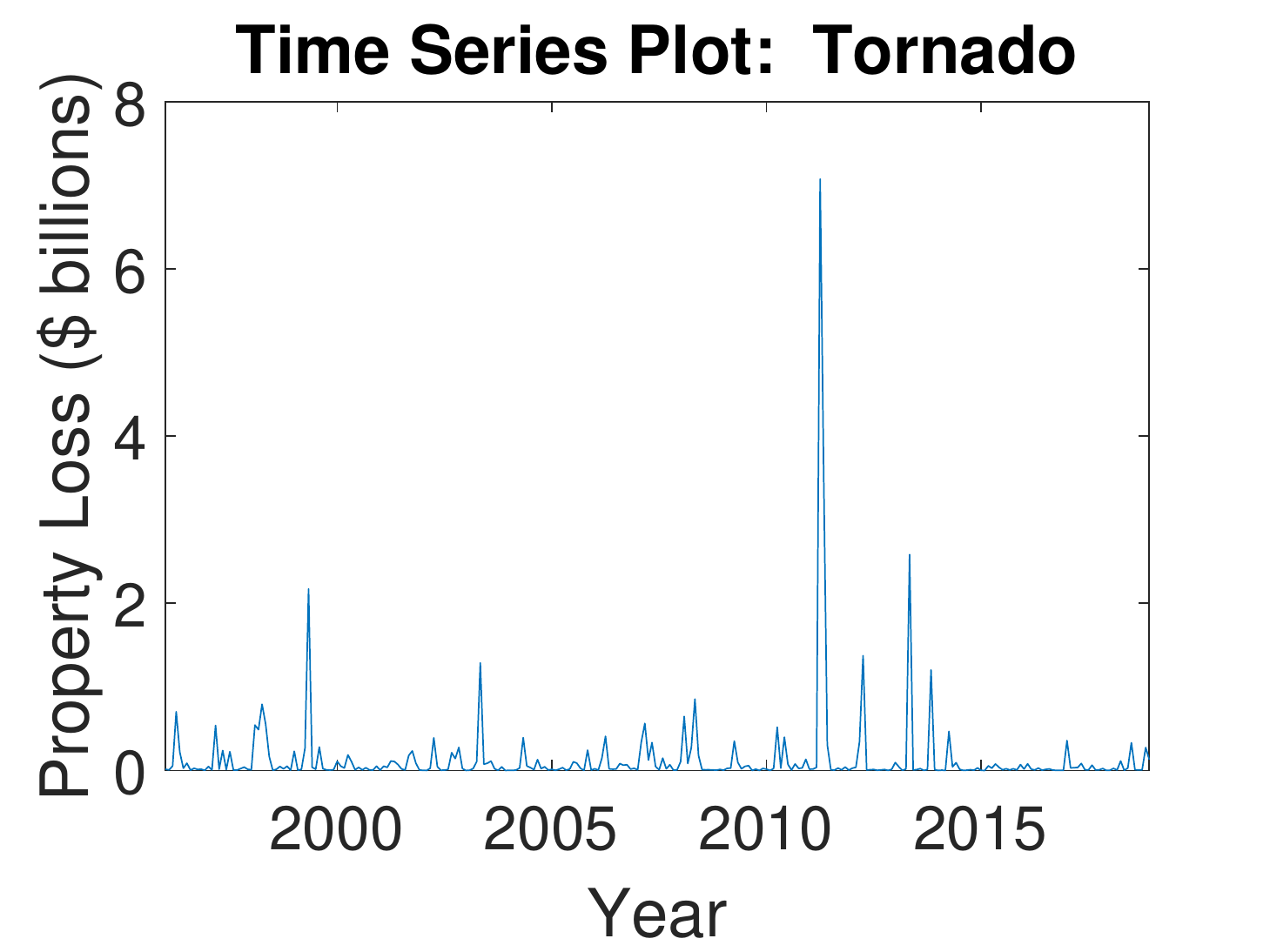}} 			
	\caption{The monthly property losses (in billions adjusted for inflation in 2019) caused by drought, flood, winter storm, thunderstorm wind, hail, and tornado events between 1996 and 2018 generated using NOAA Storm Data \citep{StormData}.}
	\label{Fig:Events}
\end{figure}

The database reports the property losses incurred by natural disasters in U.S. dollars of the given year \citep{StormDataPrep}. 
For this study, we estimate them in U.S. dollars adjusted for inflation in 2019.
Figure \ref{Fig:Events} provides examples of natural disasters between 1996 and 2018 that exemplify eccentric property losses (adjusted for inflation in 2019).\\


\begin{figure}[h!]
	\centering
	\includegraphics[width=0.6\textwidth]{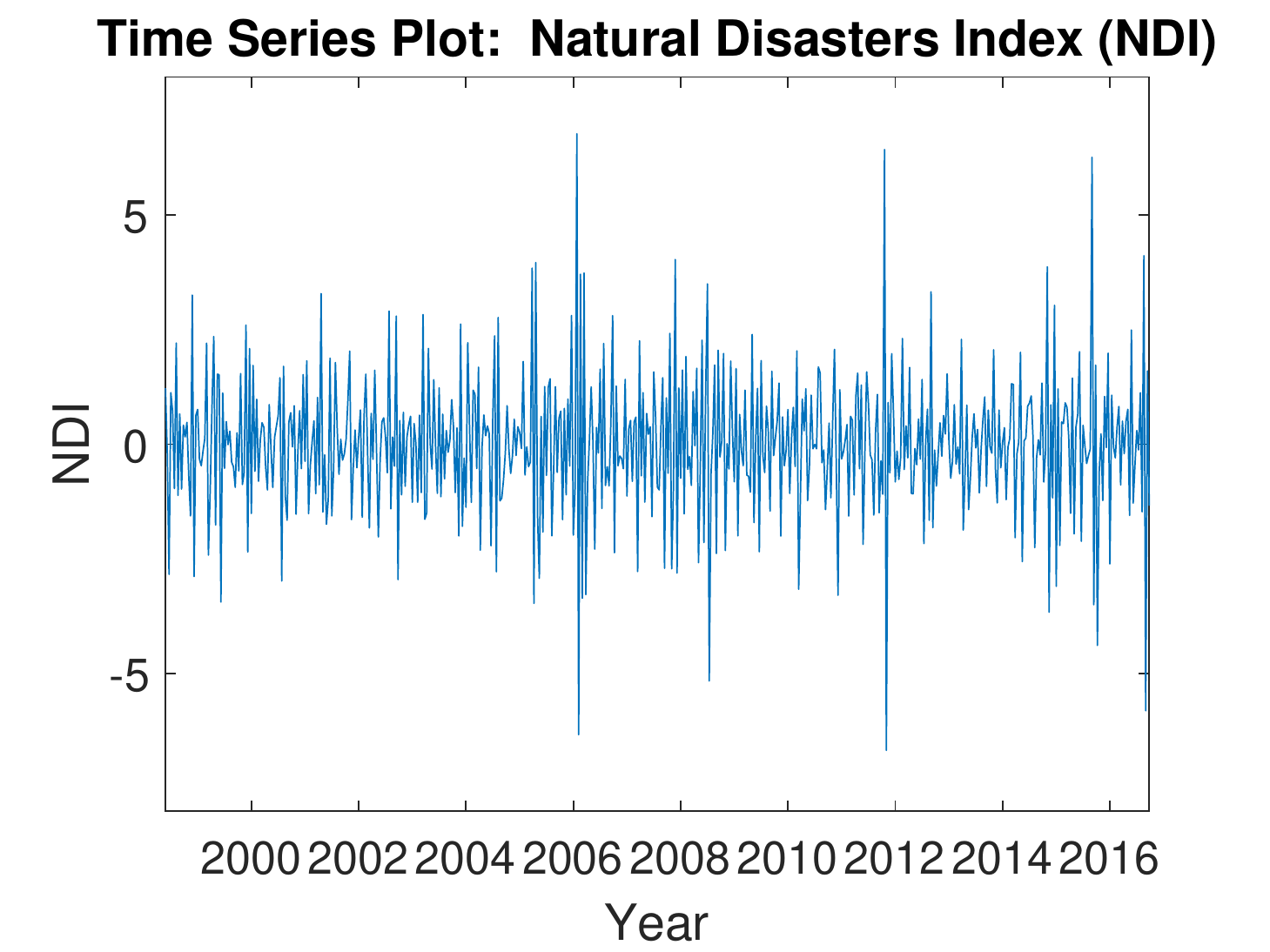}
	\caption{Our proposed Natural Disasters Index (NDI) for the United States. This NDI (\ref{Eq:Return}) is constructed using the property losses of natural disasters reported in NOAA Storm Data \citep{StormData} between 1996 and 2018.}
	\label{Fig:LossReturns_ts}
\end{figure}

\noindent \textbf{Natural Disasters Index (NDI)}\\
To obtain an equally spaced time series, we examine the cumulative property losses for all 50 types of natural disasters in two-week increments between 1996 and 2018. We define $L_t$ as the total property loss at the $t$\textsuperscript{th} biweekly period. Then, we transform this time series {$L_t$} to a stationary time series by taking the first difference (lag-1 difference) of $L_t^{0.1}$ (\ref{Eq:Return}), see Figure \ref{Fig:LossReturns_ts}. Thus, we propose a Natural Disasters Index (NDI) as follows:
\begin{equation} \label{Eq:Return}
NDI_t = L_t^{0.1}-L_{t-1}^{0.1} ,  \;\;\;\;\;\;\;\;\;\;\;\;\;\;    t = 1, \cdots, T = 552.
\end{equation}


\begin{figure}[h!]
	\subfigure[]{\includegraphics[width=0.5\textwidth]{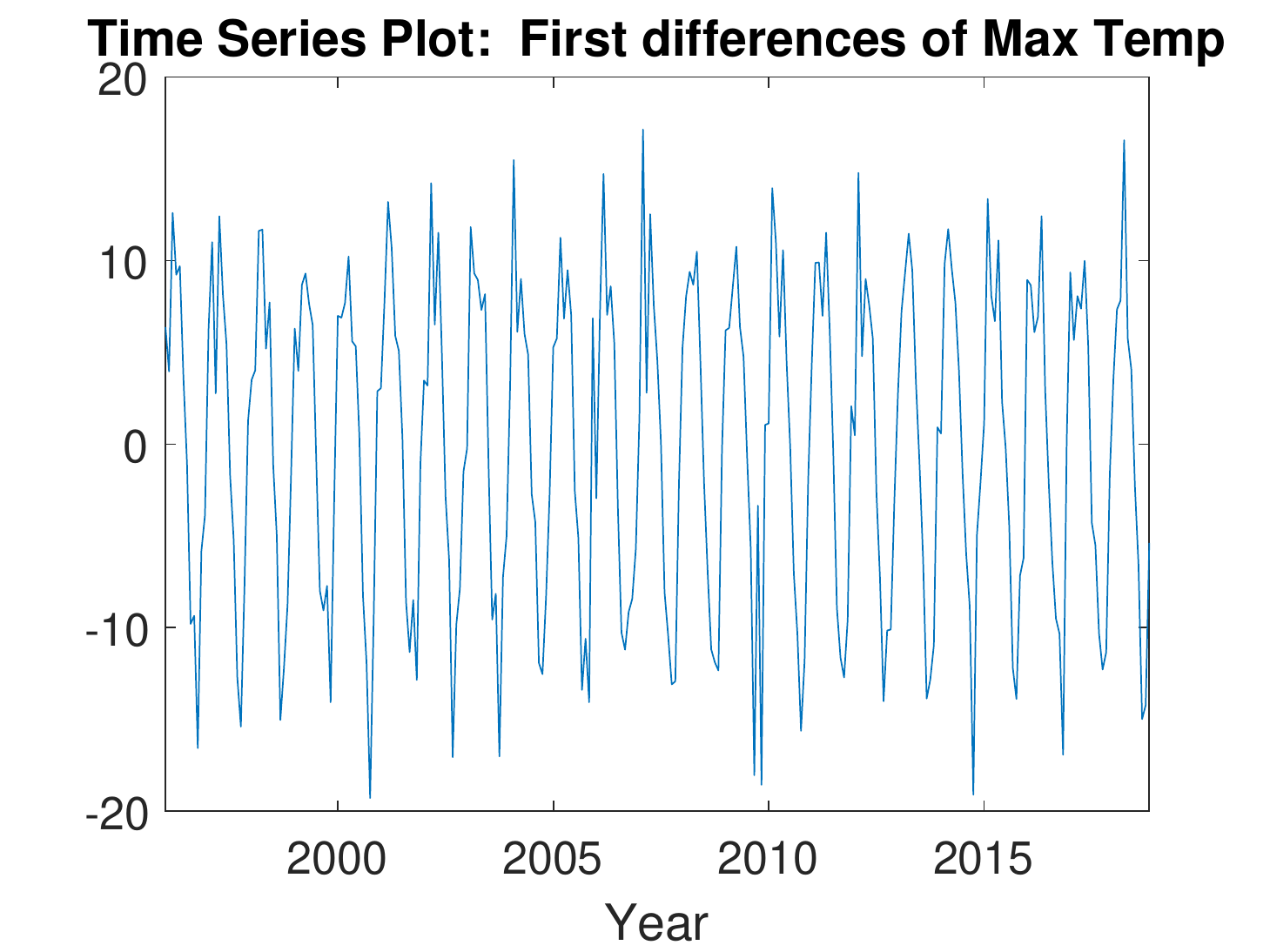}} 
	\subfigure[]{\includegraphics[width=0.5\textwidth]{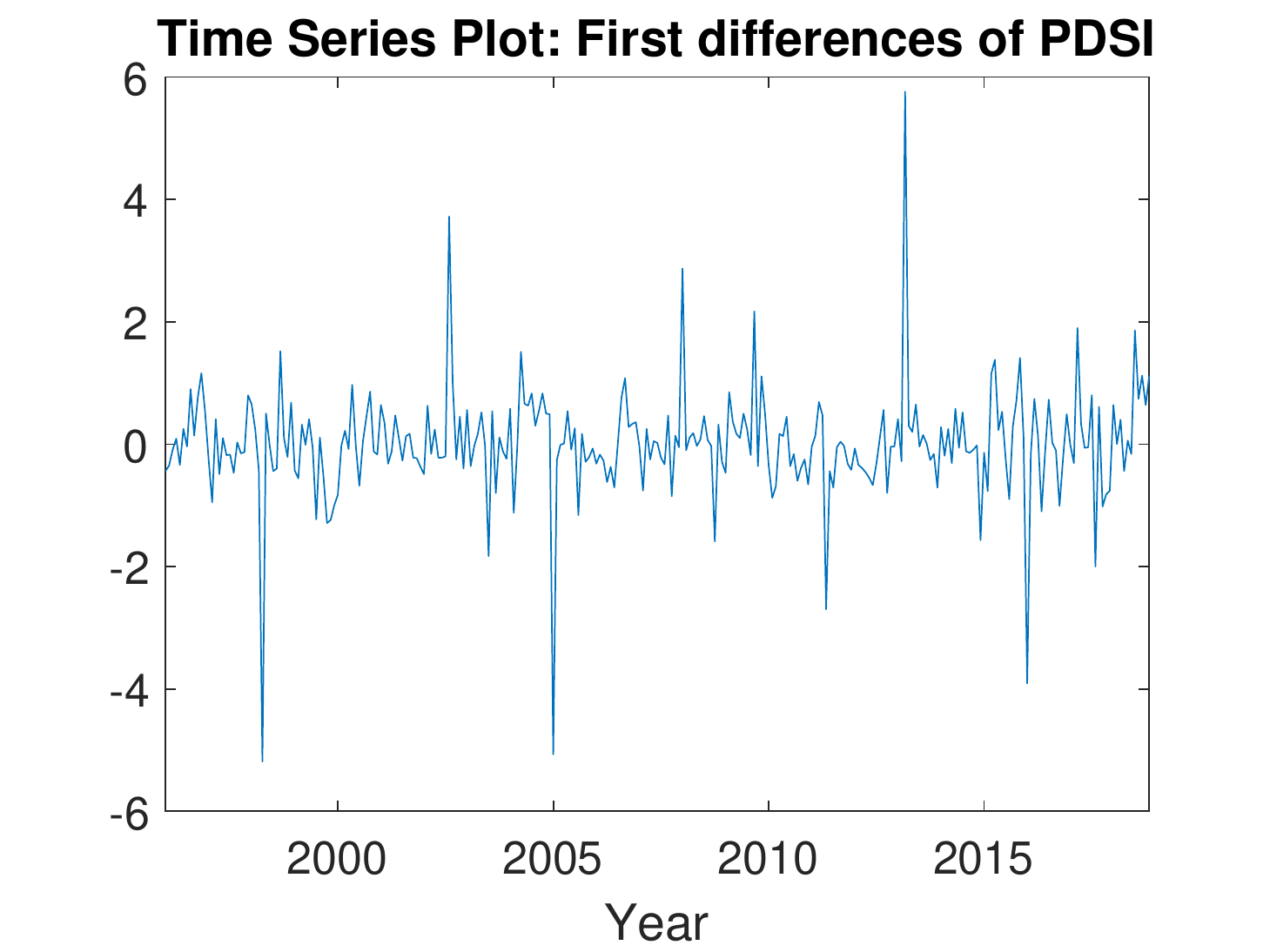}} 
	\caption{The first differences of the stress testing variables, (a) Maximum Temperature (Max Temp) and (b) Palmer Drought Severity Index (PDSI), yield stationary time series, generated using \cite{StressData} between 1996 and 2018.}
	\label{Fig:StressData_ts}
\end{figure}

For stress testing in section \ref{sec:ST}, we utilize monthly maximum temperatures and the Palmer Drought Severity Index (PDSI) used in the U.S. Climate Extremes Index (CEI) \citep{CEI,palmer1965meteorological,gleason2008revised}.
We define the reported highest temperature for each month in the U.S. as the monthly maximum temperature (measured in Fahrenheit) \citep{menne2009us,vose2014improved}.
PDSI is a measurement of severity of drought in a region for a given period \citep{heim2002review,alley1984palmer}. We use the monthly PDSI in the U.S. that assigns a value in [-4,4] on a decreasing degree of dryness (i.e., the extremely dry condition and extremely wet condition provides -4 and 4, respectively) \citep{heddinghaus1991review}. Figure \ref{Fig:StressData_ts} depicts that the first differences of both stress testing variables yield stationary time series.

\section{NDI Option Prices} \label{sec:OP}

Standard insurance and reinsurance systems encounter difficulties in reimbursing the extremely high losses caused by natural disasters.
Insurance companies seek more reliable approaches for hedging and transferring these types of intensive risks to capital market investors. 
Catastrophe risk bonds (CAT bonds) are one of the most important types of Insurance-Linked-Securities used to accomplish this.
Our proposed NDI is intended to assess the degree of future systemic risk caused by natural disasters. 
Therefore, we determine a proper model for pricing the NDI options in this section.

Options can be used for hedging, speculating, and gauging risk.
The Black-Scholes model, binomial option pricing model, trinomial tree, Monte Carlo simulation, and finite difference model are the conventional methods in option pricing. Recently, the discrete stochastic volatility based model was introduced to compute option prices and explain some well-known mispricing phenomena.
Furthermore, \cite{Duan:1995} proposes the application of discrete-time Generalized AutoRegressive Conditional Heteroskedasticity (GARCH) to price options. We extend his work by considering the standard GARCH model with Generalized Hyperbolic (GH) innovations to compute the fair values of the NDI options. We assume that the dynamic returns (\ref{Eq:Return}) follow the process 
\begin{equation}
\label{Dynamic_procss_WBI}
R_t=\log{\frac{NDI_t}{NDI_{t-1}}}=r'_t+\lambda_0 \sqrt{a_t}-\frac{1}{2}a_t+\sqrt{a_t} \epsilon_t,
\end{equation}
where $r'_t$ and $\epsilon_t$ are the risk-less rate of return and standardized residual during the time period $t$, respectively, $\lambda_0$ denotes the risk premium for the NDI, and $a_t$ is the conditional variance of returns ($R_t$) given the information set consisting of all linear functions of the past returns available during the time period $t-1$ ($F_{t-1}$), i.e., $a_t=var\left( R_t \mid F_{t-1}\right)$. We use the standard GARCH(1,1) to model 
\begin{equation}
\label{GARCH_model}
a^2_t=m+a\, a^{2}_{t-1}+b\, \epsilon_{t-1}^{2},
\end{equation}
where $m$ (constant), $a$, and $b$ are non-negative parameters of the model; each of these variables is to be estimated from the data. We assume the standardized residuals ($\epsilon_{t}$) are independent and identically distributed $GH\left(\lambda,\alpha,\beta,\delta,\mu \right)$. 
According to \cite{Blaesild:1981}, $R_t$ for given $F_{t-1}$ is distributed on real world probability space ($\mathbb{P})$ as
\begin{equation}
\label{WBI_distribution}
R_t \sim GH\left(\lambda,\frac{\alpha}{\sqrt{a_t}}, \;\frac{\beta}{\sqrt{a_t}}, \;\delta\sqrt{a_t}, \;r'_t+m_t+\mu \sqrt{a_t} \right), \;\;\;\; m_t=\lambda_0 \sqrt{a_t}-\frac{1}{2}a_t .
\end{equation}

The Esscher transformation given in \cite{Gerber:1994} is the conventional method of identifying an equivalent martingale measure to obtain a consistent price for options. Using the Esscher transformation, \cite{Chorro:2012} found that $R_t$ for given $F_{t-1}$ is distributed on the risk-neutral probability ($\mathbb{Q}$) as follows: 
\begin{equation}
\label{WBI_Q_distribution}
R_t \sim GH\left(\lambda,\frac{\alpha}{\sqrt{a_t}},\frac{\beta}{\sqrt{a_t}}+\theta_t,\delta \sqrt{a_t},r'_t+m_t+\mu \sqrt{a_t} \right),
\end{equation}
where $\theta_t$ is the solution to $MGF\left(1+\theta_t \right)= MGF\left(\theta_t \right) \, e^{r'_t}$, and $MGF$ is the conditional moment generating function of $R_{t+1}$ given $F_{t}$.

We generate future values of the NDI to price its call and put options using the Monte Carlo simulations \citep{Chorro:2012} as follows:

\begin{enumerate}
	\item Fitting GARCH(1,1) with Normal Inverse Gaussian (NIG) innovations to $L_t^{0.1}$ and forecasting $a_1^2$ (we set $t=1$).   
	\item Beginning from $t=2$, repeat the steps (a)-(c) for $t=3,4,...,T$, where $T$ is time to maturity of the NDI call option.
	\begin{enumerate}
		\item Estimating the parameter $\theta_t$ using $MGF\left(1+\theta_t \right)= MGF\left(\theta_t \right) \, e^{r'_t}$, where $MGF$ is the conditional moment generating function of $R_{t+1}$ given $F_{t}$ on $\mathbb{P}$. 
		\item Finding an equivalent distribution function for $\epsilon_t$ on $\mathbb{Q}$ and generate the value of $\epsilon_{t+1}$ under the assumption $\epsilon_{t} \sim GH(\lambda,\alpha,\beta+\sqrt{a_t}\theta_t,\delta,\mu)$ on $\mathbb{Q}$.  
		\item Computing the values of $R_{t+1}$ and $a_{t+1}$ using \eqref{Dynamic_procss_WBI} and \eqref{GARCH_model}.
		
	\end{enumerate}
	\item Generating future values of $L_t^{0.1}$ for $t=1,....,T$ on $\mathbb{Q}$ where $T$ is the time to maturity. Recursively, future values of the NDI  is obtained by 
	\begin{equation}
	NDI_{t}=R_{t}^{10}+NDI_{t-1}.
	\end{equation}
	\item Repeating steps 2 and 3 for 10,000 ($N$) times to simulate $N$ paths to compute future values of the NDI. 
\end{enumerate} 	

Then, the Monte Carlo averages approximate future values of the NDI at time $t$ for a given strike price $K$ to price its call and put options ($\hat{C}$ and $\hat{P}$, respectively)
\begin{equation} \label{Eq:Call}
\hat{C}\left(t,T,K \right)=\frac{1}{N}\,e^{-r'_t(T-t)}\sum_{i=1}^{N}\left(NDI^{(i)}_T-K \right)_{+}\ ,\\ 
\end{equation}
\begin{equation} \label{Eq:Put}
\hat{P}\left(t,T,K \right)=\frac{1}{N}\,e^{-r'_t(T-t)}\sum_{i=1}^{N}\left(K-NDI^{(i)}_T \right)_{+}. 
\end{equation}

\begin{figure}[h!]
	\centering
	\includegraphics[width=0.95\textwidth]{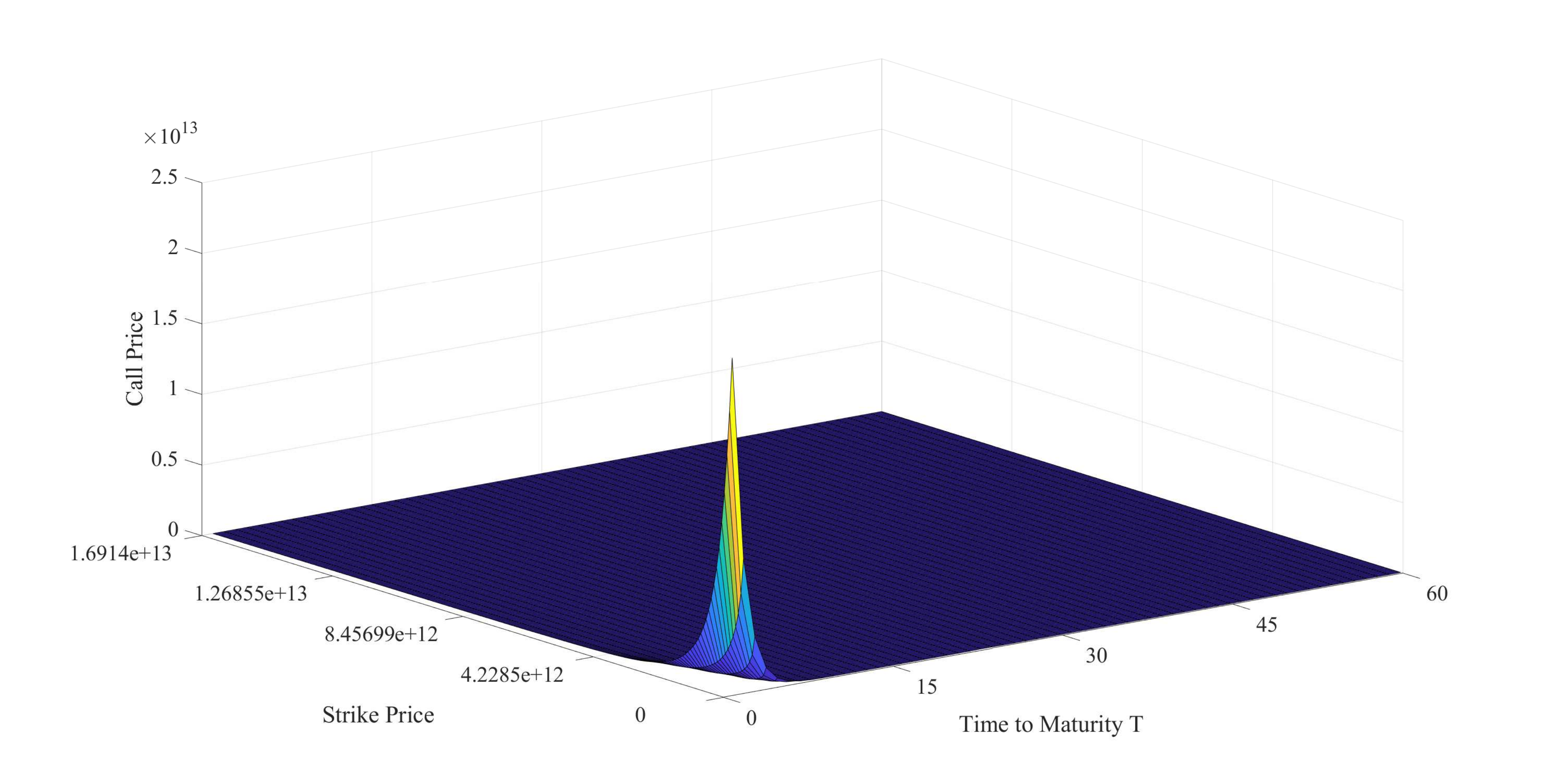}
	\caption{The call option prices (\ref{Eq:Call}) for the Natural Disasters Index (NDI) at time $t$ for a given strike price $K$ using a GARCH(1,1) model with generalized hyperbolic innovations. The Monte Carlo simulations are generated using NOAA Storm Data \citep{StormData} between 1996 and 2018.}
	\label{fig:call}
\end{figure}

\begin{figure}[h!]
	\centering
	\includegraphics[width=1\textwidth]{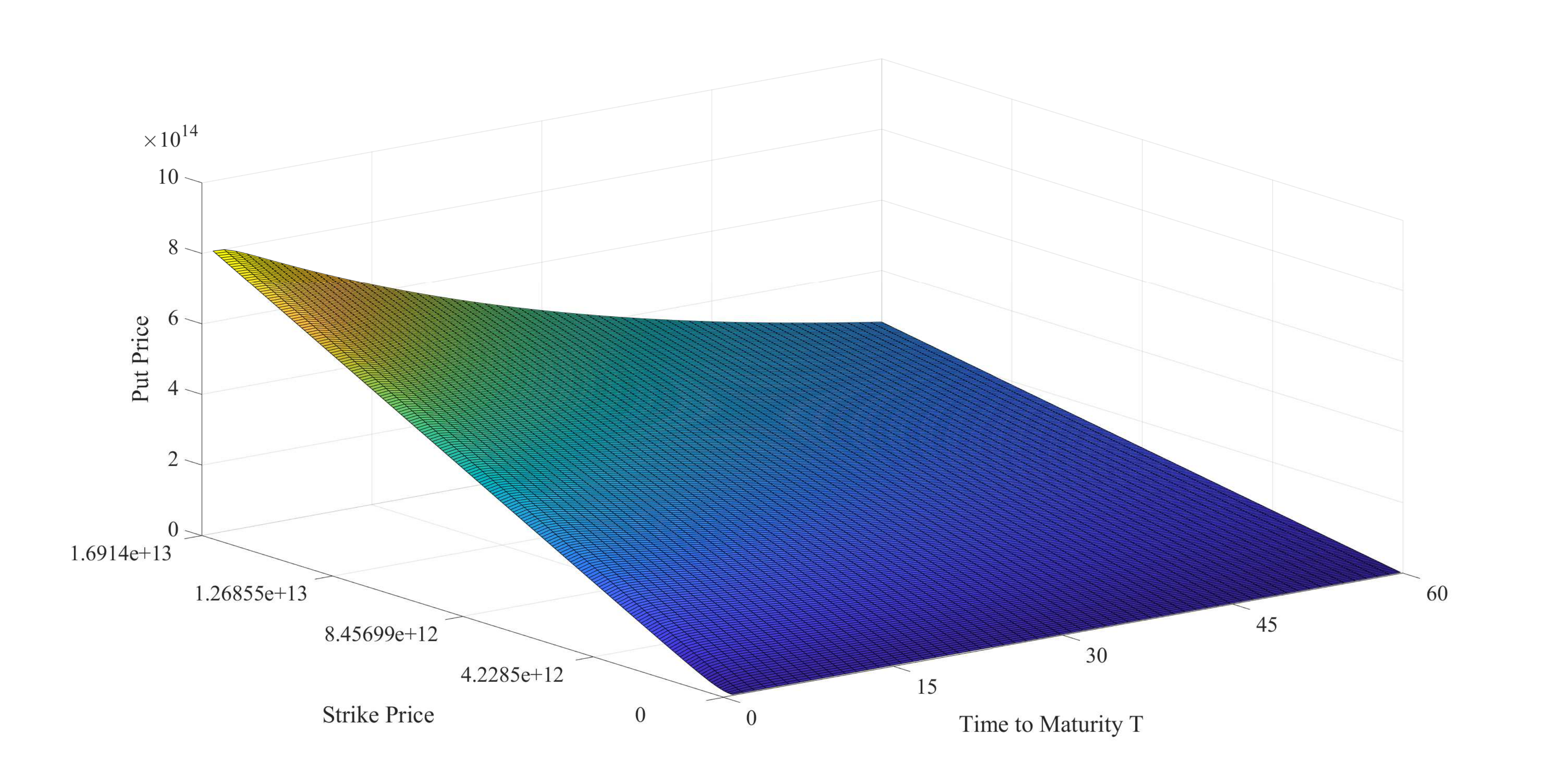}
	\caption{The put option prices (\ref{Eq:Put}) for the Natural Disasters Index (NDI) at time $t$ for a given strike price $K$ using a GARCH(1,1) model with generalized hyperbolic innovations. The Monte Carlo simulations are generated using NOAA Storm Data \citep{StormData} between 1996 and 2018.}
	\label{fig:put}
\end{figure}

\begin{figure}[h!]
	\centering
	\includegraphics[width=1\textwidth]{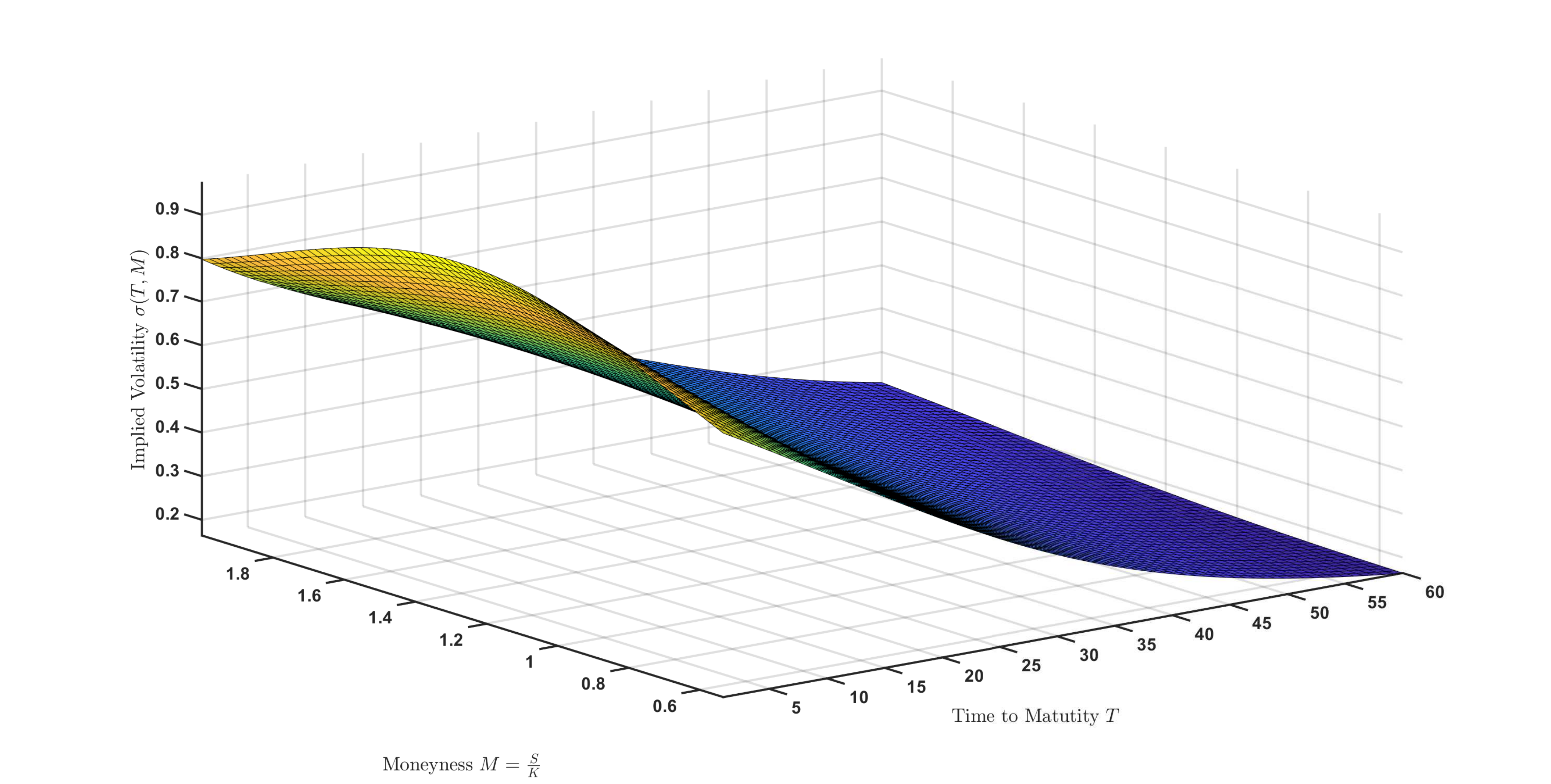}
	\caption{The Natural Disasters Index (NDI) implied volatilities against time to maturity ($T$) and moneyness ($M=S/K$, where $S$ and $K$ the stock and strike prices, respectively) using a GARCH(1,1) model with generalized hyperbolic innovations. The Monte Carlo simulations are generated using NOAA Storm Data \citep{StormData} between 1996 and 2018.}
	\label{fig:implied}
\end{figure}

We provide call and put option prices for the NDI ($\hat{C}$ and $\hat{P}$) at time $t$ for a given strike price $K$ in Figure \ref{fig:call} and \ref{fig:put}, respectively. 
These figures illustrate the relationship between time to maturity ($T$), the strike price ($K$), and option prices.
As we expected, in Figure \ref{fig:put} the put option price for NDI ($\hat{P}$) increases as the strike price increases. 
However, the call option price for NDI ($\hat{C}$) increases as the strike price decreases, see Figure \ref{fig:call}. 
Figure \ref{fig:implied} depicts the implied volatility surface against the time to maturity ($T$) and moneyness ($M=S/K$), where $S$ is the stock price. 
The observed volatility surface has an inverted volatility smile which is usually seen in periods of high market stress. 
Options with lower strike prices have higher implied volatilities compared to those with higher strike prices. 
The highest implied volatilities of options are observed in $(1.2,1.4)$ of moneyness. 
The implied volatilities tend to converge to a constant as the time to maturity converges to $60$.


\begin{figure}[h!]
	\centering
	\includegraphics[width=1\textwidth]{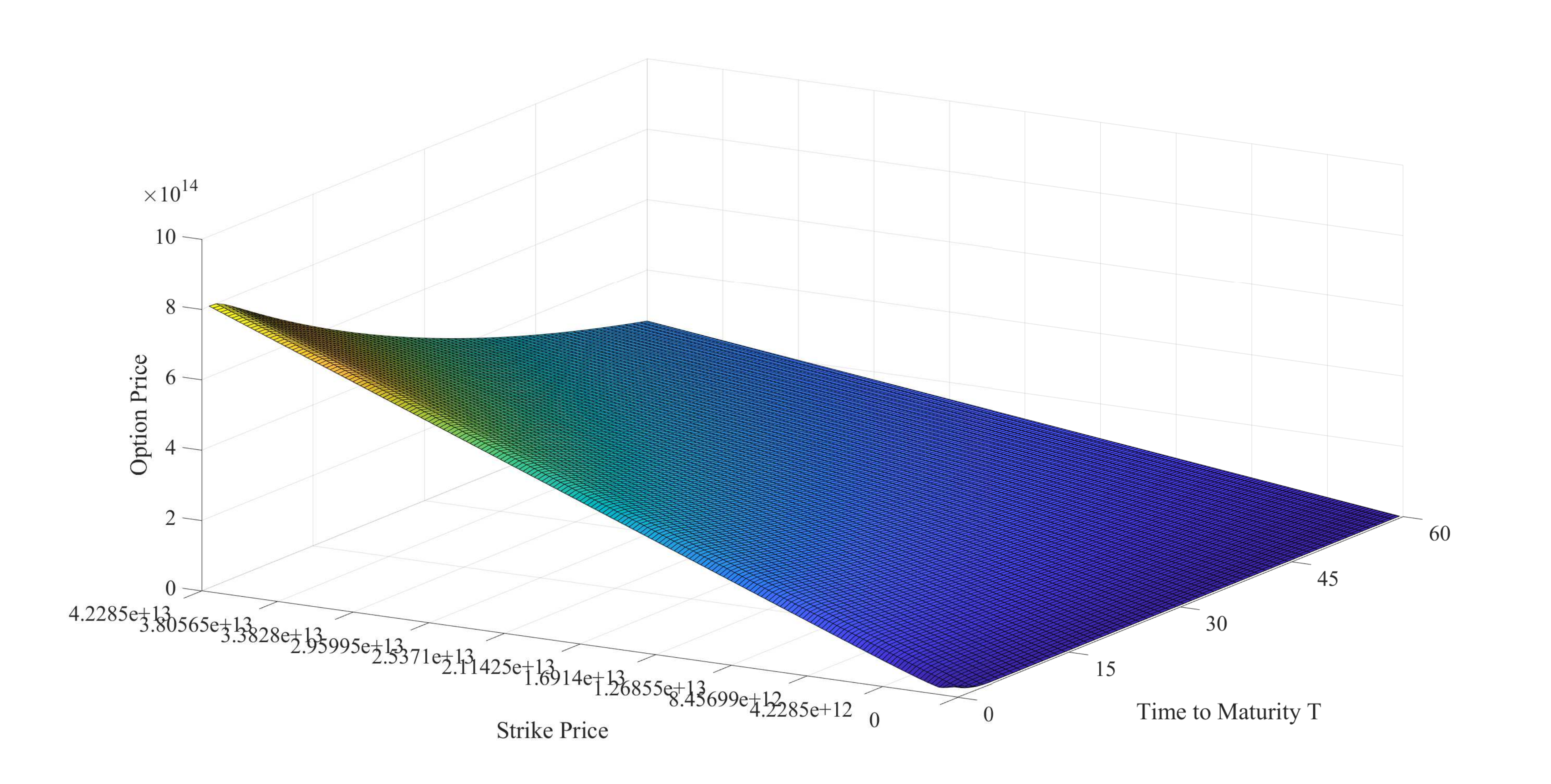}
	\caption{The call and put option prices for the Natural Disasters Index (NDI) at time $t$ for a given strike price $K$ using a GARCH(1,1) model with generalized hyperbolic innovations. The Monte Carlo simulations are generated using NOAA Storm Data \citep{StormData} between 1996 and 2018.}
	\label{fig:call-put}
\end{figure}


\section{NDI Risk Budgets} \label{sec:RB}

The risk budgets help investors as they provide the risk contributions of each component in the portfolio to the aggregate portfolio risk. To accomplish this, an investor should determine the relationships among various factors. Then, the investor can envision the amount of risk exposure (as partly) depending on the behavior of each component position. The primary strategies of assessing the center risk and tail risk contributions are portfolio standard deviation (Std), Value at Risk (VaR), and Expected Tail Loss (ETL) budgets. Some recent research applied Std and VaR for portfolio risk budgeting \footnote{See \cite{Chow:2001}, \cite{Litterman:1996}, \cite{Maillard:2010}, and \cite{Peterson:2008}.} and ETL budgets are used in \cite{Boudt:2013}. As Std and ETL are coherent risk measures, we use them as the investment strategies for our equal-weighted portfolio.

We delineate the marginal risk and risk contribution of each asset in the portfolio. We define a risk measure, $R(.)$, on the portfolio weight vector, $\textbf{w}=(w_1,w_2,...,w_n)$ where $w_i=\frac{1}{n}$ ($R(w):\mathbb{R}^n \rightarrow \mathbb{R}$). Then, the marginal contribution to risk (MCTR) of the $i$\textsuperscript{th} asset to the total portfolio risk is 
\begin{equation}
\label{MCTR}
\textit{MCTR}_i(\textbf{w}) = w_i \frac{\partial R(\textbf{w})}{\partial w_i}.
\end{equation}
The MCTR of the $k$\textsuperscript{th} subset is 
\begin{equation}
\label{PCTR}
\textit{MCTR}_{M_k}(\textbf{w}) = \sum_{i \in M_k}^{}\textit{MCTR}_i (\textbf{w}),
\end{equation}
where $M_k\,\, \subseteq \,\left\lbrace1,2,...,n \right\rbrace$ denote $s$ subsets of portfolio assets. The percent contribution to risk (PCTR) of the $i$\textsuperscript{th} asset to the total portfolio risk is 

\begin{equation}
\label{PC_to_R}
\textit{PCTR}_i(\textbf{w}) =\frac{\textit{MCTR}_i(\textbf{w})}{\sum_{i=1}^{n}\textit{MCTR}_i(\textbf{w})}.
\end{equation}

\begin{figure}[h!]
	\centering
	\includegraphics[width=1\textwidth]{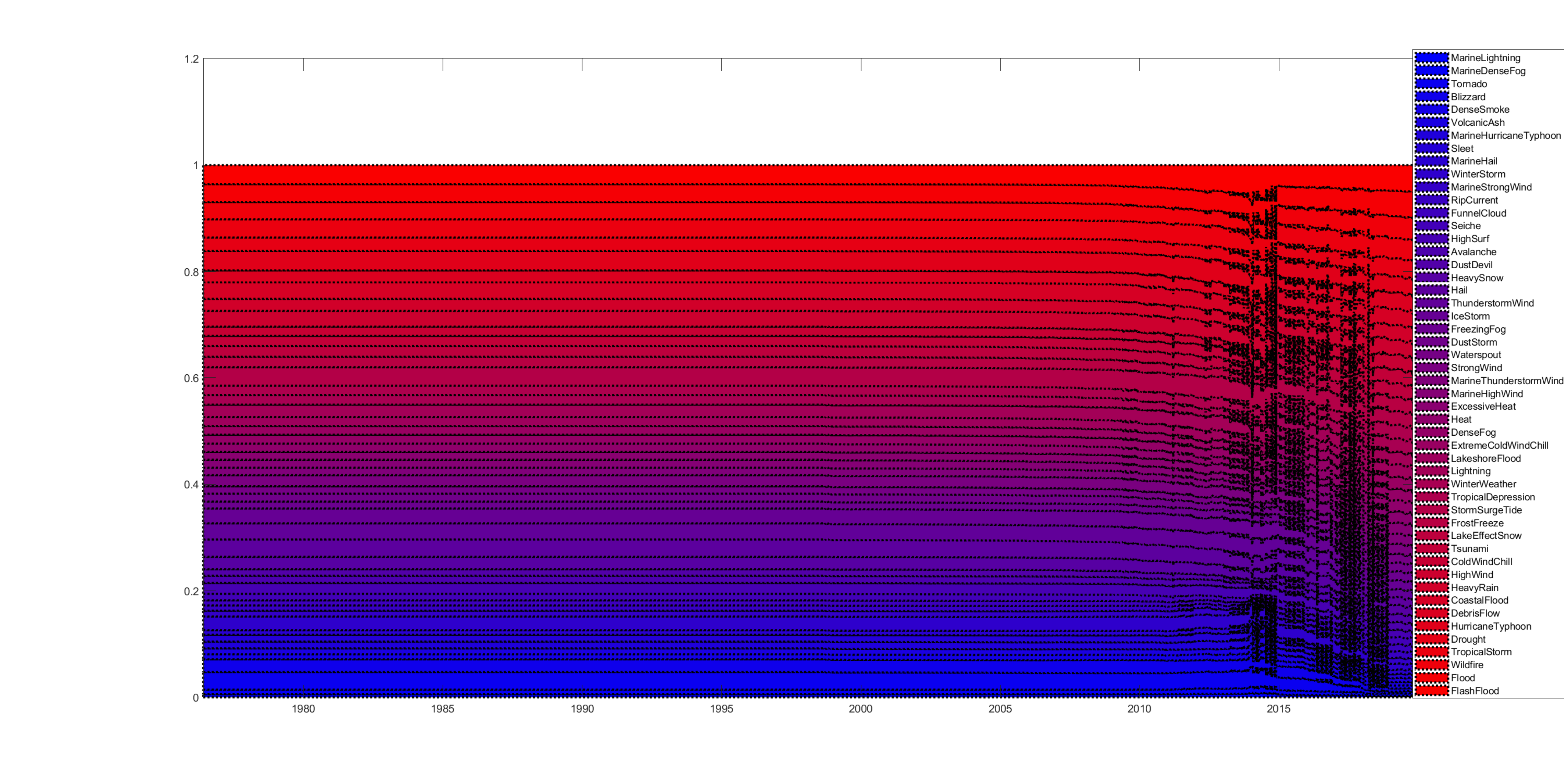}
	\caption{The percent contribution to risk (PCTR) of the expected tail loss (ETL) risk budgets for the Natural Disasters Index (NDI) at 95\% level. The legend depicts the severe weather events in ascending order of their PCTR of ETL risk budgets at 95\% level. The results are generated from NOAA Storm Data \citep{StormData} between 1996 and 2018.}
	\label{fig:RiskBudgeting_PCTR95}
\end{figure}

\begin{figure}[h!]
	\centering
	\includegraphics[width=1\textwidth]{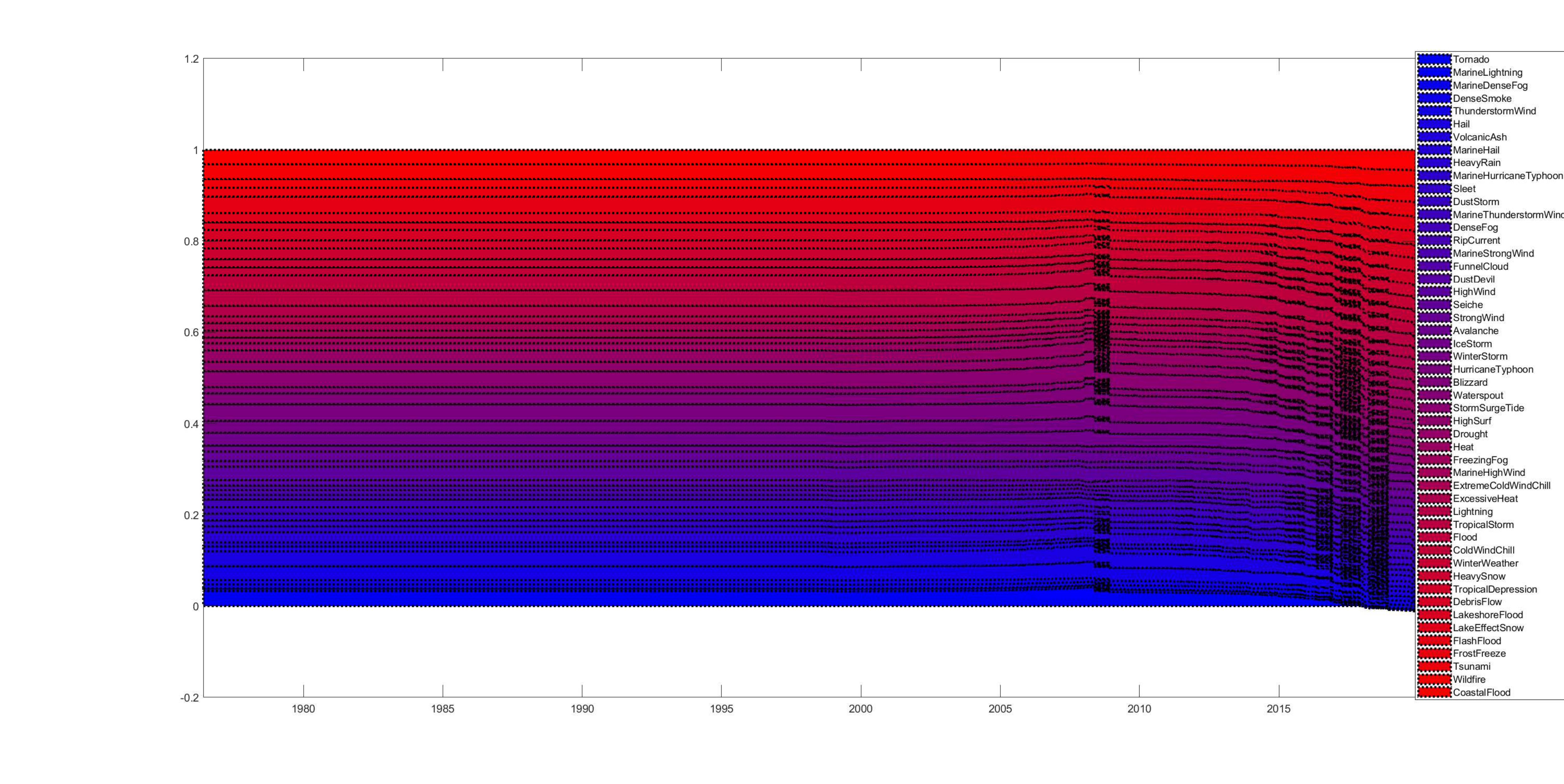}
	\caption{The percent contribution to risk (PCTR) of the expected tail loss (ETL) risk budgets for the Natural Disasters Index (NDI) at 99\% level. The legend depicts the severe weather events in ascending order of their PCTR of ETL risk budgets at 99\% level. The results are generated from NOAA Storm Data \citep{StormData} between 1996 and 2018.}
	\label{fig:RiskBudgeting_PCTR99}
\end{figure}

\begin{figure}[h!]
	\centering
	\includegraphics[width=1\textwidth]{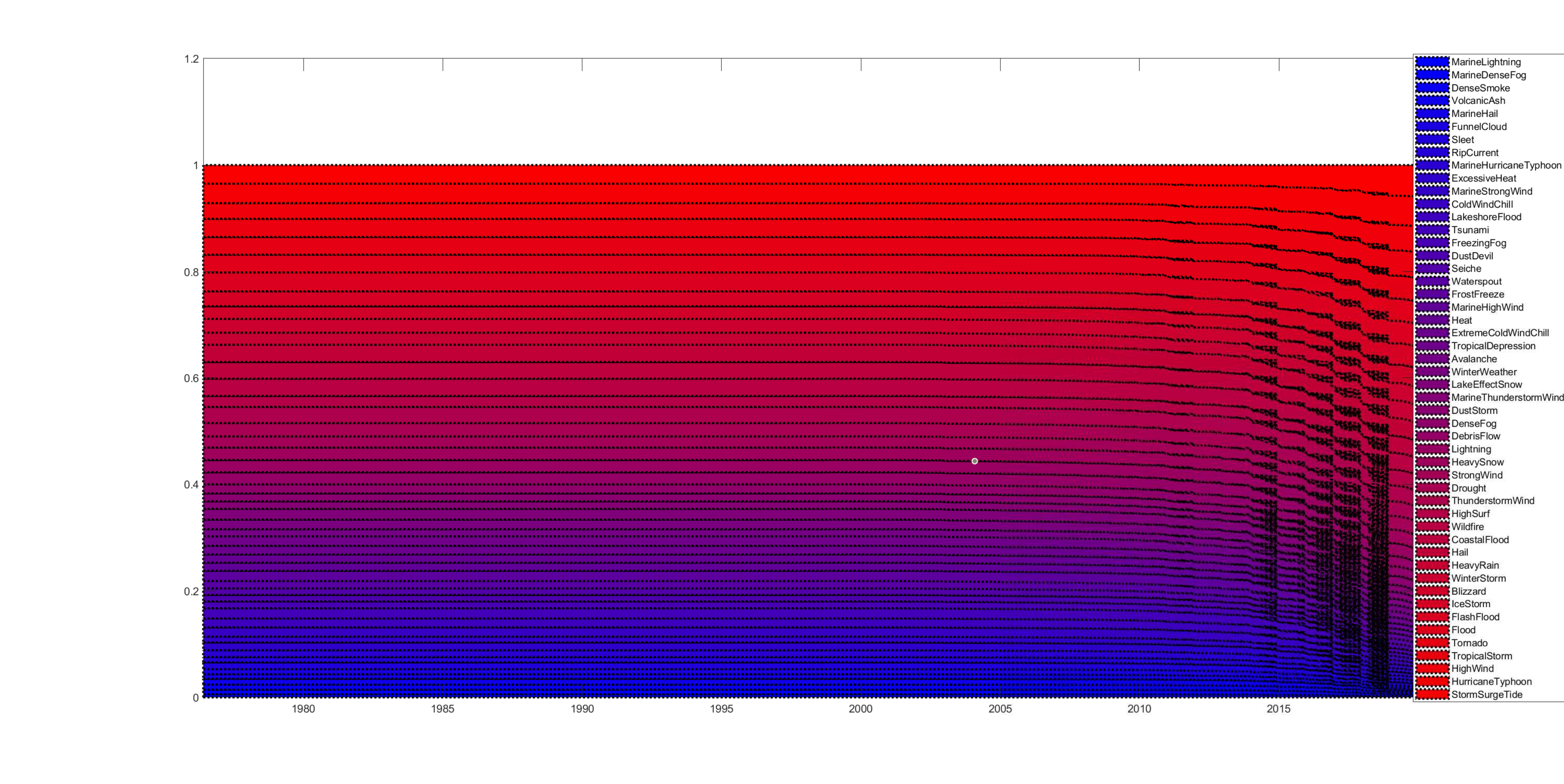}
	\caption{The percent contribution to risk (PCTR) of the standard deviation (Std) risk budgets for the Natural Disasters Index (NDI). The legend depicts the severe weather events in ascending order of their PCTR of Std risk budgets. The results are generated from NOAA Storm Data \citep{StormData} between 1996 and 2018.}
	\label{fig:RiskBudgeting_PCTRSTD}
\end{figure}

Since a large number of observations are involved in our analysis, we use a rolling-method for risk budgeting. We use the first 400 data (biweekly loss returns) at each window as in-sample-data and the last 400 data as out-of-sample data. The results of the rolling method for finding risk contributions across the time are depicted in Figures \ref{fig:RiskBudgeting_PCTR95}-\ref{fig:RiskBudgeting_PCTRSTD}. 
We calculate Std and ETL for risk contributions in our portfolio. Table \ref{tab:risk_budget} reports the estimated risk allocations within the equal-weighted portfolio. According to the results, the main center risk contributors are tornado, tropical storm, flood, ice storm, and flash flood. However, flash flood, flood, and wildfire are the main tail risk contributors at $95\%$ level. Thus, flash flood and flood are the main risk contributors in our portfolio. Among center risk contributors, tornado is one of the tail risk diversifiers in our portfolio with a negative tail risk contribution at $99\%$ level. 



\begin{longtable}{@{}lcccccc@{}}
	\toprule
	Severe Weather Event &
	\begin{tabular}[c]{@{}c@{}}MCTR \\ ETL (95)\end{tabular} & \begin{tabular}[c]{@{}c@{}}PCTR \\ ETL (95)\end{tabular} & \begin{tabular}[c]{@{}c@{}}MCTR \\ ETL (99)\end{tabular} & \begin{tabular}[c]{@{}c@{}}PCTR \\ ETL (99)\end{tabular} & \begin{tabular}[c]{@{}c@{}}MCTR\\  Std\end{tabular} & \begin{tabular}[c]{@{}c@{}}PCTR \\ Std\end{tabular} \\ \midrule
	Marine Lightning & 0.0002 & 0.01\% &0.0004 &0.02\% &0.0365 &0.21\%                                              \\
	Marine Dense Fog & 0.0003 & 0.02\% & 0.0006 & 0.03\%  & 0.0552  & 0.32\%                                              \\
	Tornado & 0.0076 & 0.49\% & -0.0219 & -1.13\% & 0.7692 & 4.49\%                                              \\
	Blizzard & 0.0076 & 0.49\% & 0.0379 & 1.96\%   & 0.6046 & 3.53\%                                              \\
	Dense Smoke & 0.0088 & 0.57\% & 0.0183 & 0.94\%  & 0.0562  & 0.33\%                                              \\
	Volcanic Ash & 0.0104 & 0.68\% & 0.0214 & 1.10\%  & 0.0624 & 0.36\%                                              \\
	Marine Hurricane Typhoon & 0.0121 & 0.78\% & 0.0241 & 1.24\%  & 0.1039   & 0.61\%                                              \\
	Sleet & 0.0122 & 0.79\%  & 0.0241 & 1.24\%  & 0.0917 & 0.54\%                                              \\
	Marine Hail & 0.0124  & 0.80\%  & 0.0226 & 1.16\% & 0.0631 & 0.37\%                                              \\
	Winter Storm & 0.0162 & 1.05\% & 0.0373 & 1.92\%  & 0.5815 & 3.40\%                                              \\
	Marine Strong Wind & 0.0166 & 1.08\% & 0.0305 & 1.57\% & 0.1070  & 0.62\%                                              \\
	Rip Current & 0.0187 & 1.21\%  & 0.0301 & 1.55\% & 0.0940  & 0.55\%                                              \\
	Funnel Cloud & 0.0202  & 1.31\% & 0.0318 & 1.64\%  & 0.0680 & 0.40\%                                              \\
	Seiche & 0.0204 & 1.32\% & 0.0359 & 1.85\% & 0.1343 & 0.78\%                                              \\
	High Surf & 0.0206 & 1.33\% & 0.0409  & 2.11\% & 0.5224 & 3.05\%                                              \\
	Avalanche & 0.0223 & 1.44\% & 0.0359 & 1.85\% & 0.1964 & 1.15\%                                              \\
	Dust Devil & 0.0233 & 1.51\% & 0.0352 & 1.82\% & 0.1287 & 0.75\%                                              \\
	Heavy Snow & 0.0243 & 1.58\% & 0.0555 & 2.86\% & 0.4441 & 2.59\%                                              \\
	Hail & 0.0244 & 1.58\% & 0.0193 & 0.99\% & 0.5757 & 3.36\%                                              \\
	Thunderstorm Wind & 0.0252 & 1.63\% & 0.0184 & 0.95\% & 0.5072 & 2.96\%                                              \\
	Ice Storm & 0.0254 & 1.64\%  & 0.0367 & 1.89\% & 0.7052 & 4.12\%                                              \\
	Freezing Fog & 0.0256 & 1.66\% & 0.0436 & 2.25\%  & 0.1156 & 0.67\%                                              \\
	Dust Storm & 0.0257 & 1.66\% & 0.0281 & 1.45\%  & 0.2943 & 1.72\%                                              \\
	Waterspout & 0.0259 & 1.68\% & 0.0396 & 2.04\% & 0.1509 & 0.88\%                                              \\
	Strong Wind & 0.0290 & 1.88\% &  0.0359 & 1.85\% & 0.4462 & 2.61\%                                              \\
	Marine Thunderstorm Wind & 0.0292 & 1.89\%  & 0.0298 & 1.54\%   & 0.2261 & 1.32\%                                              \\
	Marine High Wind & 0.0321 & 2.08\% & 0.0439 & 2.26\% & 0.1578 & 0.92\%                                              \\
	Excessive Heat & 0.0325 & 2.11\% & 0.0486 & 2.51\%  & 0.1057 & 0.62\%                                              \\
	Heat & 0.0325 & 2.11\% & 0.0428  & 2.20\% & 0.1624 & 0.95\%                                              \\
	Dense Fog & 0.0337 & 2.19\% & 0.0298 & 1.54\% & 0.3104 & 1.81\%                                              \\
	Extreme Cold Wind Chill & 0.0350 & 2.27\% & 0.0462 & 2.38\%  & 0.1814 & 1.06\%                                              \\
	Lakeshore Flood & 0.0359 & 2.33\%  & 0.0577 & 2.98\% & 0.1130 & 0.66\%                                              \\
	Lightning & 0.0361 & 2.34\% & 0.0492 & 2.54\% & 0.4109 & 2.40\%                                              \\
	Winter Weather & 0.0384 & 2.49\% & 0.0554 & 2.85\% & 0.1976 & 1.15\%                                              \\
	Tropical Depression & 0.0397 & 2.58\% & 0.0567 & 2.92\%  & 0.1816 & 1.06\%                                              \\
	Storm Surge Tide & 0.0398 & 2.58\% & 0.0402 & 2.07\% & 1.0165 & 5.94\%                                              \\
	Frost Freeze & 0.0403 & 2.61\% & 0.0630 & 3.25\% & 0.1570 & 0.92\%                                              \\
	Lake Effect Snow & 0.0405 & 2.63\% & 0.0589 & 3.04\%  & 0.2124 & 1.24\%                                              \\
	Tsunami & 0.0406 & 2.63\% & 0.0678 & 3.49\% & 0.1143 & 0.67\%                                              \\
	Cold Wind Chill & 0.0428 & 2.77\% & 0.0549 & 2.83\% & 0.1072  & 0.63\%                                              \\
	High Wind & 0.0437  & 2.83\% & 0.0355 & 1.83\% & 0.8385 & 4.90\%                                              \\
	Heavy Rain & 0.0439 & 2.84\% & 0.0240 & 1.24\% & 0.5813 & 3.39\%                                              \\
	Coastal Flood & 0.0455 & 2.95\% & 0.0877 & 4.52\% & 0.5504 & 3.21\%                                              \\
	Debris Flow & 0.0486 & 3.15\% & 0.0568 & 2.93\%  & 0.3216 & 1.88\%                                              \\
	Hurricane Typhoon & 0.0486  & 3.15\% & 0.0376 & 1.94\% & 0.9643 & 5.63\%                                              \\
	Drought & 0.0524 & 3.40\% & 0.0416 & 2.14\% & 0.4786 & 2.79\%                                              \\
	Tropical Storm & 0.0598 & 3.88\% & 0.0509 & 2.62\%  & 0.8305 & 4.85\%                                              \\
	Wildfire & 0.0639 & 4.14\% & 0.0678 & 3.50\% & 0.5399 & 3.15\%                                              \\
	Flood & 0.0752 & 4.87\% & 0.0512 & 2.64\% & 0.7292 & 4.26\%                                              \\
	Flash Flood & 0.0766 & 4.96\% & 0.0598 & 3.08\% & 0.7210 & 4.21\%                                              \\ \bottomrule
	\caption{The standard deviation (Std) and expected tail loss (ETL) (at 95\% and 99\% levels) risk budgets for the Natural Disasters Index (NDI). Marginal contribution to risk and the percent contribution to risk are given by MCTR (\ref{MCTR}) and PCTR (\ref{PCTR}), respectively.}
	\label{tab:risk_budget}
\end{longtable}


\section{Stress Testing Analysis for the NDI} \label{sec:ST}

In finance, stress testing is an analysis intended to determine the strength of a financial instrument and its resilience to the economic crisis. Stress testing is a form of scenario analysis used by regulators to investigate the robustness of a financial instrument is in inevitable crashes. In risk management, this helps to determine portfolio risks and serves as a tool for hedging strategies required to mitigate against potential losses.

In this section, we assess the performance of the NDI via stress testing using monthly maximum temperature (Max Temp) and the Palmer Drought Severity Index (PDSI) as stressors (refer to section \ref{sec:data}). 
Instead of working with each factor, we use the first differences of them as returns that yield stationary time series, see Figure \ref{Fig:StressData_ts}.

The two series of returns inherit serial correlation and dependence according to the results of Ljung-Box test results (p-values are less than 0.05). 
Thus, to capture linear and nonlinear dependencies in data sets, we put the series through the ARMA(1,1)-GARCH(1,1) filter with Student-t innovations. Then, we consider the sample innovations obtained from the aforementioned filter for our analysis.

\begin{figure}[h!]
	\subfigure{\includegraphics[width=0.5\textwidth]{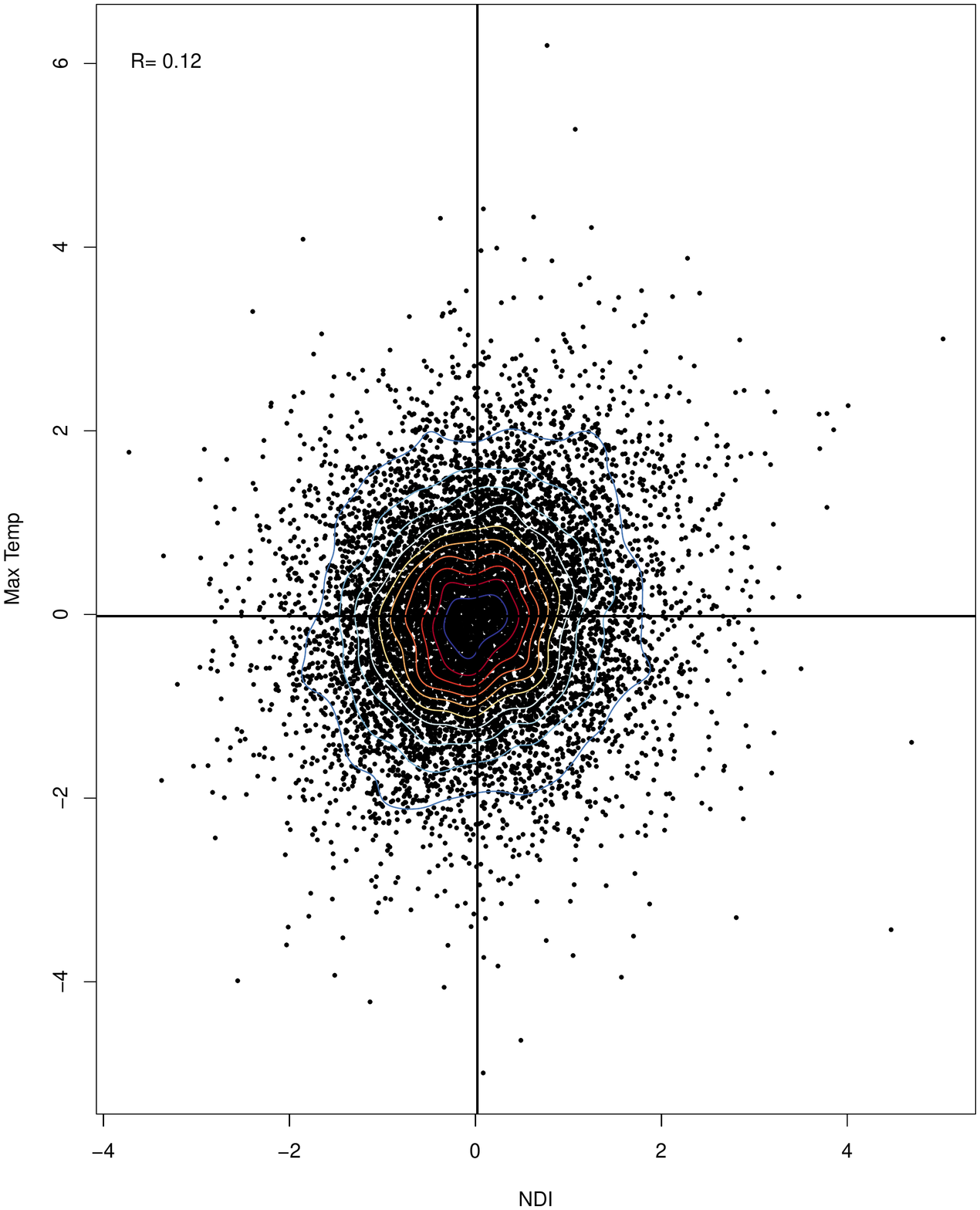}} 
	\subfigure{\includegraphics[width=0.5\textwidth]{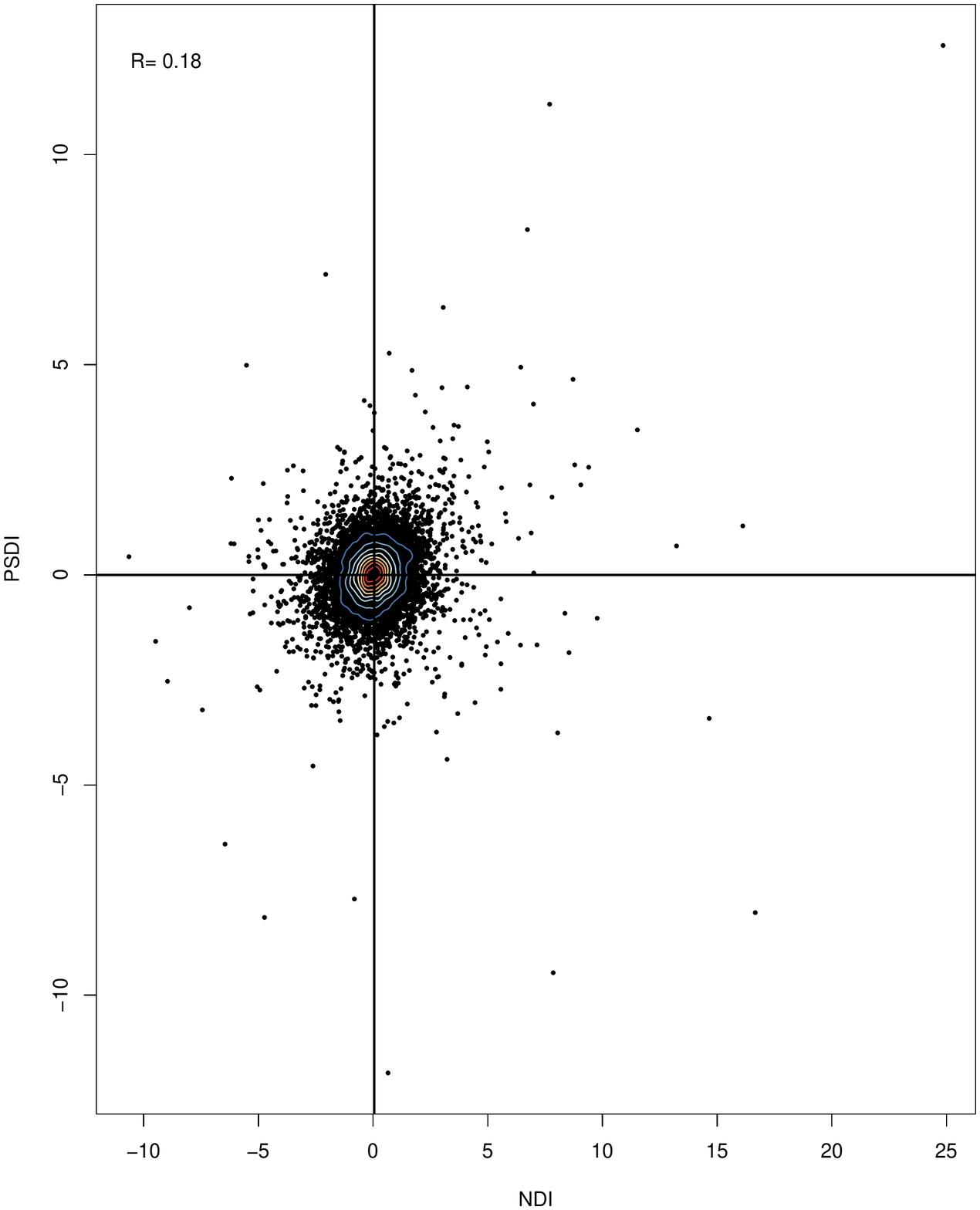}} 
	\caption{The generated joint densities of the returns of monthly maximum temperature (Max Temp) and the Natural Disasters Index (NDI), and the Palmer Drought Severity Index (PDSI) and the NDI (right panel) using the fitted bivariate NIG models of the joint distributions of independent and identically distributed standardized residuals. The figures depict the simulated values and the contour plots of the joint densities.}
	\label{Fig:aa}
\end{figure}

We fit bivariate NIG models to the joint distributions of independent and identically distributed standardized residuals of each factor and the NDI (Total Loss): Max Temp vs NDI and PSDI vs NDI. 
Then, we simulate 10,000 values from the models of factors to perform the scenario analysis and to compute the systemic risk measures. 
Figure \ref{Fig:aa} shows the fitted contour plots from each model, overlaid with the 10,000 simulated values. The empirical correlation coefficients based on the observed data suggest a weak positive relationship between the factors and the NDI $(R\simeq 0.155)$.


There are various measures of systemic risk used to assess the impact of negative events on the stress factors. \cite{adrian2011covar}  proposed Conditional Value-at-Risk (CoVaR): the change in the value at risk of the financial system conditional on an institution being under distress relative to its median state. \cite{girardi2013systemic} improved the definition of financial distress from an institution being exactly at its VaR to being at most at its VaR (CoVaR). As the CoVaR is a coherent risk measure \cite[see][]{acerbi2002expected}, changing VaR to CoVaR allows us to consider more severe distress events, back-test CoVaR, and improve its monotonicity concerning the dependence parameter. The definition of CoVaR by \cite{girardi2013systemic} was based on the conditional distribution of a random variable $Y$ given a stress event for a random variable $X$. \cite{Mainik14} defined an alternative CoVaR notion in terms of copulas. They showed that conditioning on $X\leq VaR_{\alpha}(X)$ improves the response to dependence between $X$ and $Y$ compared to conditioning on $X= VaR_{\alpha}(X)$. Therefore, we use the variant of CoVaR developed by \cite{Mainik14} for our study.

We define the distributions of $Y$ and $X$ are given by $F_{Y}$ and $F_{X}$, respectively, and $F_{Y|X}$ is the conditional distribution of $Y$ given $X$. Then, CoVaR
at level $q$, $\text{CoVaR}_q$ (or $\xi_{q}$), is defined as 
\begin{equation}\label{eq:def-CoVaR}
\xi_q := \text{CoVaR}_q := F^{-1}_{Y|X\leq F^{-1}_{X}(q)}
\left( q \right) = \text{VaR}_{q}\left(Y|X\leq\text{VaR}_{q}(X)\right),
\end{equation}
where $\text{VaR}_{q}\left(X\right)$ denotes the VaR of $X$ at level $q$, which is same as the $q$\textsuperscript{th} quantile of $X$ ($F^{-1}_{X}(q)$). In \cite{Mainik14}, CoVar for the closely associated Expected Shortfall (ES) is defined as the tail mean beyond VaR:
\begin{equation}\label{eq:def-CoES}
\text{CoES}_q := \E\left(Y|Y\leq\xi_{q},X\leq\text{VaR}_{q}(X)\right).
\end{equation}
Furthermore, \cite{ZariCOETL} has proposed
\begin{equation}\label{eq:def-CoES}
\text{CoETL}_q := \E\left(Y|Y\leq\text{VaR}_{q}(Y),X\leq\text{VaR}_{q}(X)\right).
\end{equation}

Table \ref{tab:risk-measures} reports the left-tail systemic risk measures on the NDI at different levels based on stressing the factors (Max Temp and PDSI).
At 5\% and 10\% stress levels, stress on Max Temp seems to have a marginally more meaningful impact on the NDI than the stress on PDSI. However, at the highest stress level (1\%), the results show stress on Max Temp has a greater significant impact on the NDI compared to that of PDSI.


\begin{table}[]
	\centering
	\begin{tabular}{@{}lcccc@{}}
		\toprule
		\multicolumn{1}{c}{\multirow{2}{*}{Stress Factors}} & \multirow{2}{*}{Stress Levels} & \multicolumn{3}{c}{Risk Measure on the NDI (Left Tail)} \\ \cmidrule(l){3-5} 
		\multicolumn{1}{c}{} & & CoVaR & CoES & CoETL \\ \cmidrule(r){1-5}
		\multirow{3}{*}{Max Temp} 
		& 10\% & -1.868 & -2.527 & -2.027\\
		& 5\% & -2.472 & -3.324 & -2.469\\
		& 1\% & -4.568 & -5.305 & -3.756\\ 	\cmidrule(r){1-5}
		\multirow{3}{*}{PDSI}
		& 10\%  & -1.348 & -2.368 & -1.551\\
		& 5\% & -2.221 & -4.001 & -2.159\\
		& 1\% & -7.863 & -17.625 & -4.265\\ \cmidrule(l){1-5} 
	\end{tabular}
	\caption{The left-tail systemic risk measures (CoVaR, CoES, and CoETL) on the Natural Disasters Index (NDI) at different stress levels based on stressing the factors monthly maximum temperature (Max Temp) and the Palmer Drought Severity Index (PDSI).}
	\label{tab:risk-measures}
\end{table}

\newpage
\section{Discussion and Conclusion} \label{sec:DC}

We proposed the Natural Disasters Index, NDI (\ref{Eq:Return}), using the United States as a model with property losses reported in NOAA Storm Data \citep{StormData} between 1996 and 2018.
In order to establish the NDI, we provided an evaluation framework using three promising approaches: (1) option pricing, (2) risk budgeting, and (3) stress testing.

We determined the fair values of the NDI options using a discrete-time GARCH model with GH innovations and then simulated Monte Carlo averages to approximate call and put option prices (\ref{Eq:Call}),(\ref{Eq:Put}).
The relationships among time to maturity, strike price, and option prices help to construct and valuate insurance-type financial instruments.
Then, we disaggregated the cumulative risk attributed to natural disasters to our equally-weighted portfolio (i.e., we investigated the risk contribution of each type of natural disaster).
The Std and ETL risk budgets for the NDI yield that flood and flash flood are the main risk contributors in our portfolio.
Finally, we assessed the performance of the NDI via a stress testing analysis using Max Temp and PDSI as stressors.
We found the stress on Max Temp significantly impacts the NDI compared to that of the PDSI at the highest stress level (1\%).

The proposed NDI is an attempt to address a financial instrument for hedging the intrinsic risk induced by the property losses caused by natural disasters in the United States.
The main objective of the NDI is to forecast the degree of future systemic risk caused by natural disasters.
This information could forewarn the insurers and corporations allowing them to transfer insurance risk to capital market investors. 
Hence the issuance of the NDI will conspicuously help to bridge the gap between the capital and insurance markets. 
While the NDI is specifically constructed for the United States, it could be modified to calculate the risk in other regions or countries using a data set comparable to NOAA Storm Data \cite{StormData}.

\clearpage
\normalem



\end{document}